\documentclass{aa}  

\usepackage[english]{babel}
\usepackage{amsmath}
\usepackage{amssymb}
\newcommand{\ignore}[1]{}
\usepackage{natbib}
\usepackage{amssymb}
\usepackage{txfonts}
\usepackage{longtable}
\usepackage{graphicx}

\newcommand{\mearth}{M_\oplus}

\def\ms{\hbox{\,m\,s$^{-1}$}}         
\def\m2s2{\hbox{\,m$^{2}$\,s$^{-2}$}} 
\def\sini{\hbox{sin\,$i$}}      

\begin{document}

   \title{The HADES RV Programme with HARPS-N@TNG \\GJ\,3998: An early M-dwarf hosting a system of Super-Earths\thanks{Based on: observations made with the Italian {\it Telescopio Nazionale Galileo} (TNG), operated on the island of La Palma by the INAF -
          {\it Fundaci{\'o}n Galileo Galilei} at the {\it Roche de Los Muchachos} Observatory of the {\it Instituto de Astrof{\'i}sica de Canarias} (IAC); photometric
	  observations made with the APACHE array located at the Astronomical Observatory of the Aosta Valley; photometric observations made with the robotic telescope
	  APT2 (within the EXORAP program) located at Serra La Nave on Mt. Etna.}       }

\titlerunning{The HADES RV Programme}

   \author{L. Affer\inst{1}
          \and
          G. Micela\inst{1}
	  \and
	  M. Damasso\inst{2}
	  \and
	  M. Perger\inst{3}
	  \and
	  I. Ribas\inst{3}
	  \and
	  A. Su{\'a}rez Mascare{\~n}o\inst{4,5}
	  \and
	  J. I. Gonz{\'a}lez Hern{\'a}ndez\inst{4,5}
	  \and
	  R. Rebolo\inst{4,5,6}
	  \and
	  E. Poretti\inst{7}
	  \and
	  J. Maldonado\inst{1}
	  \and
	  G. Leto\inst{8}
	  \and
	  I. Pagano\inst{8}
	  \and
	  G. Scandariato\inst{8}
	  \and
	  R. Zanmar Sanchez\inst{8}
          \and
	  A. Sozzetti\inst{2}
	  \and
	  A. S. Bonomo\inst{2}
	  \and
	  L. Malavolta\inst{9,10}
	  \and 
	  J. C. Morales\inst{3}
	  \and
	  A. Rosich\inst{3}
	  \and
	  A. Bignamini\inst{11}
	  \and
	  R. Gratton\inst{10}
	  \and
	  S. Velasco\inst{4,5}
	  \and
	  D. Cenadelli\inst{2,12}
	  \and
	  R. Claudi\inst{10}
	  \and
	  R. Cosentino\inst{13}
	  \and
	  S. Desidera\inst{10}
	  \and
	  P. Giacobbe\inst{2}
	  \and
	  E. Herrero\inst{3}
	  \and
	  M. Lafarga\inst{3}
	  \and
	  A. F. Lanza\inst{8}
	  \and
	  E. Molinari\inst{13}
	  \and
	  G. Piotto\inst{9,10}
	  }
 
   \institute{INAF - Osservatorio Astronomico di Palermo, Piazza del Parlamento 1, 90134 Palermo, Italy\\
              \email{affer@astropa.inaf.it}
	      \and
	      INAF - Osservatorio Astrofisico di Torino, via Osservatorio 20, 10025 Pino Torinese, Italy
	      \and 
	      Institut de Ciències de l'Espai (CSIC-IEEC), Campus UAB, Carrer de Can Magrans s/n, 08193 Cerdanyola del Vall{\'e}s, Spain
	      \and
	      Instituto de Astrof{\'{i}}sica de Canarias, 38205 La Laguna, Tenerife, Spain
	      \and
              Universidad de La Laguna, Dpto. Astrof{\'{i}}sica, 38206 La Laguna, Tenerife, Spain
              \and
	      Consejo Superior de Investigaciones Cient{\'{i}}ficas, 28006, Madrid, Spain
	      \and
              INAF - Osservatorio Astronomico di Brera, via E. Bianchi 46, 23807 Merate (LC), Italy
	      \and
	      INAF - Osservatorio Astrofisico di Catania, via S. Sofia 78, 95123 Catania, Italy
	      \and
	      Dipartimento di Fisica e Astronomia G. Galilei, Universit{\'a} di Padova, Vicolo dell'Osservatorio 2, 35122, Padova, Italy
	      \and
	      INAF - Osservatorio Astronomico di Padova, Vicolo dell'Osservatorio 5, 35122, Padova, Italy 
	      \and
	      INAF - Osservatorio Astronomico di Trieste, via Tiepolo 11, 34143, Trieste, Italy
	      \and
	      Osservatorio Astronomico della Regione Autonoma Valle d'Aosta, Fraz. Lignan 39, 11020, Nus (Aosta), Italy
	      \and
	      INAF - Fundaci{\'o}n Galileo Galilei, Rambla Jos{\'e} Ana Fernandez P{\'e}rez 7, 38712, Bre{\~n}a Baja, TF, Spain
              }


 
  \abstract
   {Many efforts to detect Earth-like planets around low-mass stars are presently devoted in almost every extra-solar planet search. M dwarfs are considered ideal targets for Doppler
   radial velocity searches because their low masses and luminosities make low-mass planets orbiting in their habitable zones more easily detectable than those around 
   higher mass stars. Nonetheless, the statistics of frequency of low-mass planets hosted by low mass stars remains poorly constrained.}
   {Our M-dwarf radial velocity monitoring with HARPS-N within the GAPS (Global architectures of Planetary Systems) -- ICE (Institut de Ciències de l'Espai/CSIC-IEEC) -- IAC (Instituto de Astrof{\'{i}}sica de Canarias)
   project\thanks{http://www.oact.inaf.it/exoit/EXO-IT/Projects/Entries/2011/12/27\_GAPS.html} can provide a major contribution to the widening of the current statistics through the in-depth analysis of accurate radial velocity observations in a narrow range of spectral sub-types (79 stars, between dM0 to dM3). Spectral accuracy will enable us to reach the precision needed to detect small planets with a few earth masses. Our survey will bring a contribute to the surveys devoted to the search for planets around M-dwarfs, mainly focused on the M-dwarf population of the northern emisphere, for which we will provide an estimate of the planet occurence. }
   {We present here a long duration radial velocity monitoring of the M1 dwarf star GJ\,3998 with HARPS-N to identify periodic signals in the data. Almost simultaneous photometric observations were carried out within the APACHE and EXORAP programs to characterize the stellar activity and to distinguish from the periodic signals those due to activity and to the presence of planetary companions. We run an MCMC simulation and use Bayesian model selection to determine the number of planets in this system, to estimate their orbital parameters and minimum masses and for a proper treatment of the activity noise.  }
   {The radial velocities have a dispersion in excess of their internal errors due to at
   least four superimposed signals, with periods of 30.7, 13.7, 42.5 and 2.65 days. Our data are well described by a 2-planet Keplerian (13.7~d and 2.65~d) and 2 sinusoidal functions (stellar activity, 30.7~d and 42.5~d) fit. The analysis of spectral indices based on Ca II H \& K and H$\alpha$ lines demonstrates
that the periods of 30.7 and 42.5 days are due to chromospheric inhomogeneities modulated by stellar rotation and differential rotation. This result is supported by photometry and is consistent with the results on differential rotation of M stars obtained with $Kepler$. The shorter periods of $13.74\pm\, 0.02\, d$ and $2.6498\pm\, 0.0008\, d$ are
   well explained with the presence of two planets, with minimum masses of $6.26^{+0.79}_{-0.76}$\, M$_\oplus$ and $2.47\pm\, 0.27$\, M$_\oplus$ and distances of 0.089 AU and 0.029 AU from the host, respectively.}
   {}


   \maketitle
%

\section{Introduction}

In the two decades since the discovery of the first giant planetary mass companion to a main sequence star \citep{may95} the search for and characterization of extrasolar planets has quickly developed to become a major field of modern-day astronomy. Thanks to concerted efforts with a variety of observational techniques (both from the ground and in space), thousands of confirmed and candidate planetary systems are known to-date (e.g., http://www.exoplanet.eu - \citealp{sch11}), encompassing orders of magnitude in mass and orbital separation. 
The frontier today is being pushed ever closer to the identification of potentially habitable small-mass planets with a well-determined rocky composition similar to Earth's. Planetary systems harboring objects with these characteristics are likely to be discovered first around primaries with later spectral types than the Sun's. 

Stars in the lower main sequence (M dwarfs) constitute the vast majority ($>70-75\%$) of all stars, both in the Solar neighbourhood and in the Milky Way as a whole \citep{henry06,win15}. They are particularly promising targets for exoplanet search programs for a number or reasons. In particular, the favourable mass and radius ratios lead to readily detectable radial-velocity (RV) and transit signals produced by terrestrial-type planets. Furthermore, the low luminosities of M dwarfs imply that the boundaries of their habitable zones (HZ) are located at short separations \citep[typically between 0.02 AU and 0.2 AU, see e.g.][]{man07}, making rocky planets orbiting within them more easily detectable with present-day observing facilities than those around more massive stars \citep[e.g.][]{cha07}. Finally, the favorable planet-star contrast ratios for small stars enable the best opportunities in the near future for detailed characterization studies of small planets and their atmospheres \citep[e.g.][]{sea10}.

\begin{figure}
\centering
\includegraphics[width=9.cm]{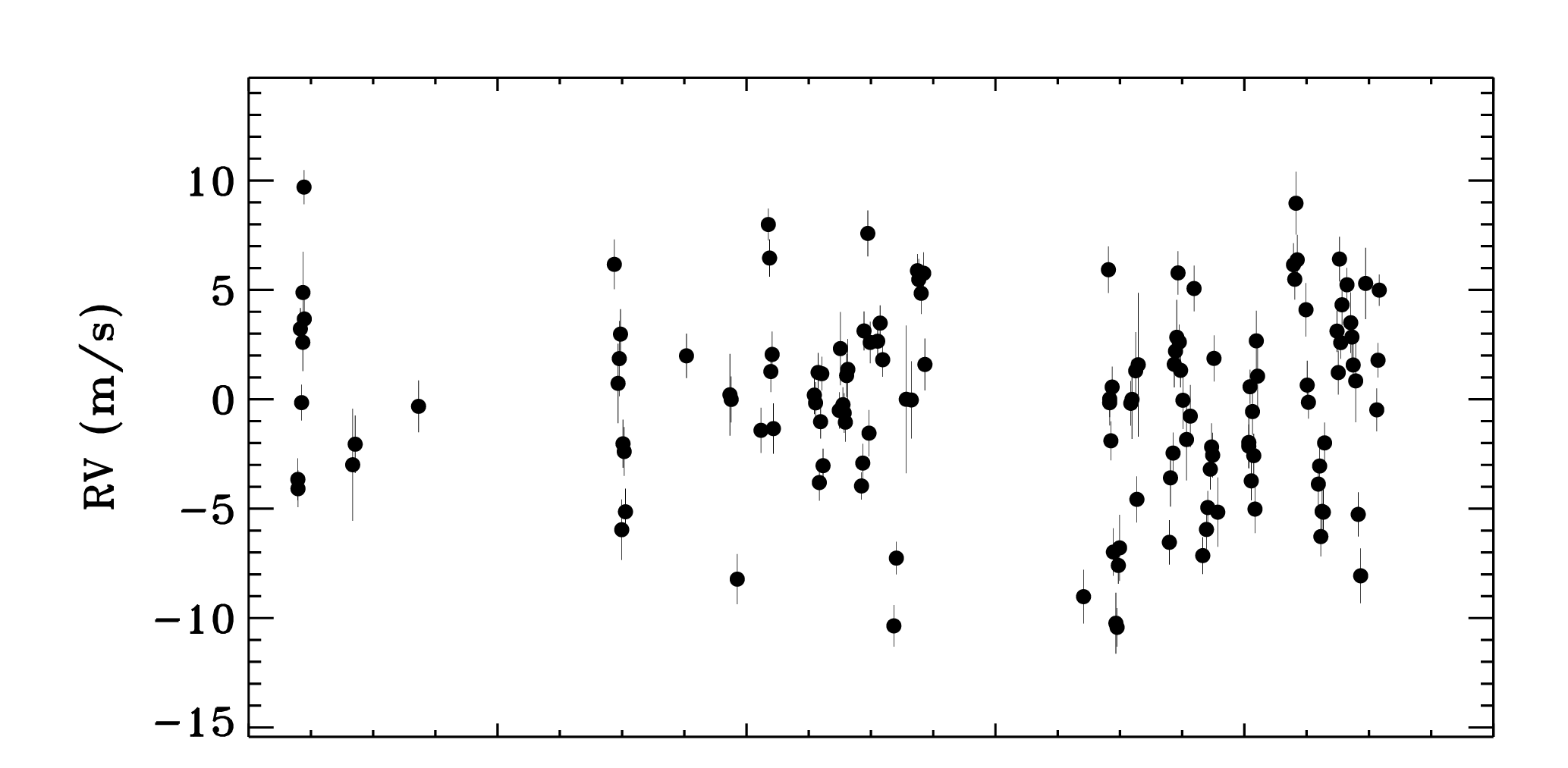}
\includegraphics[width=9.cm]{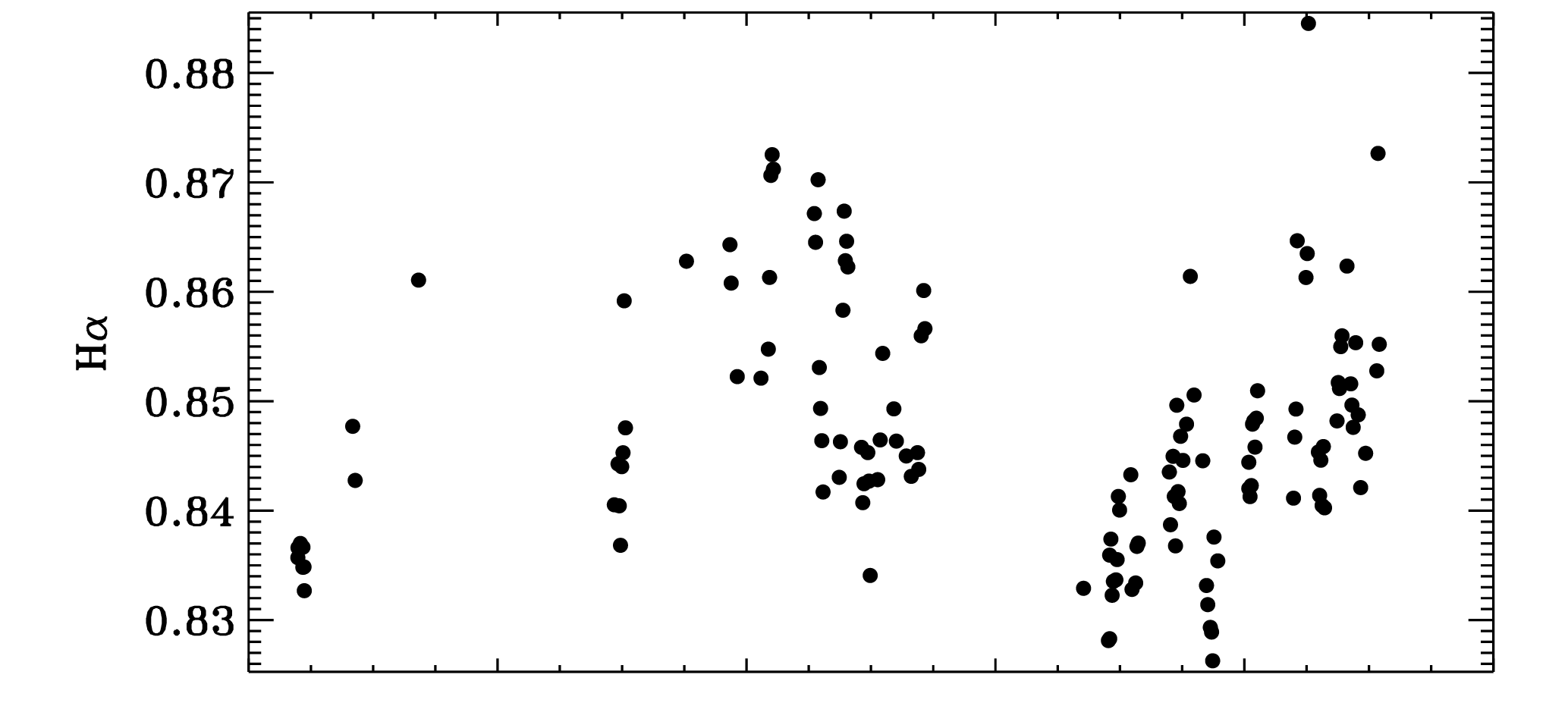}
\includegraphics[width=9.cm]{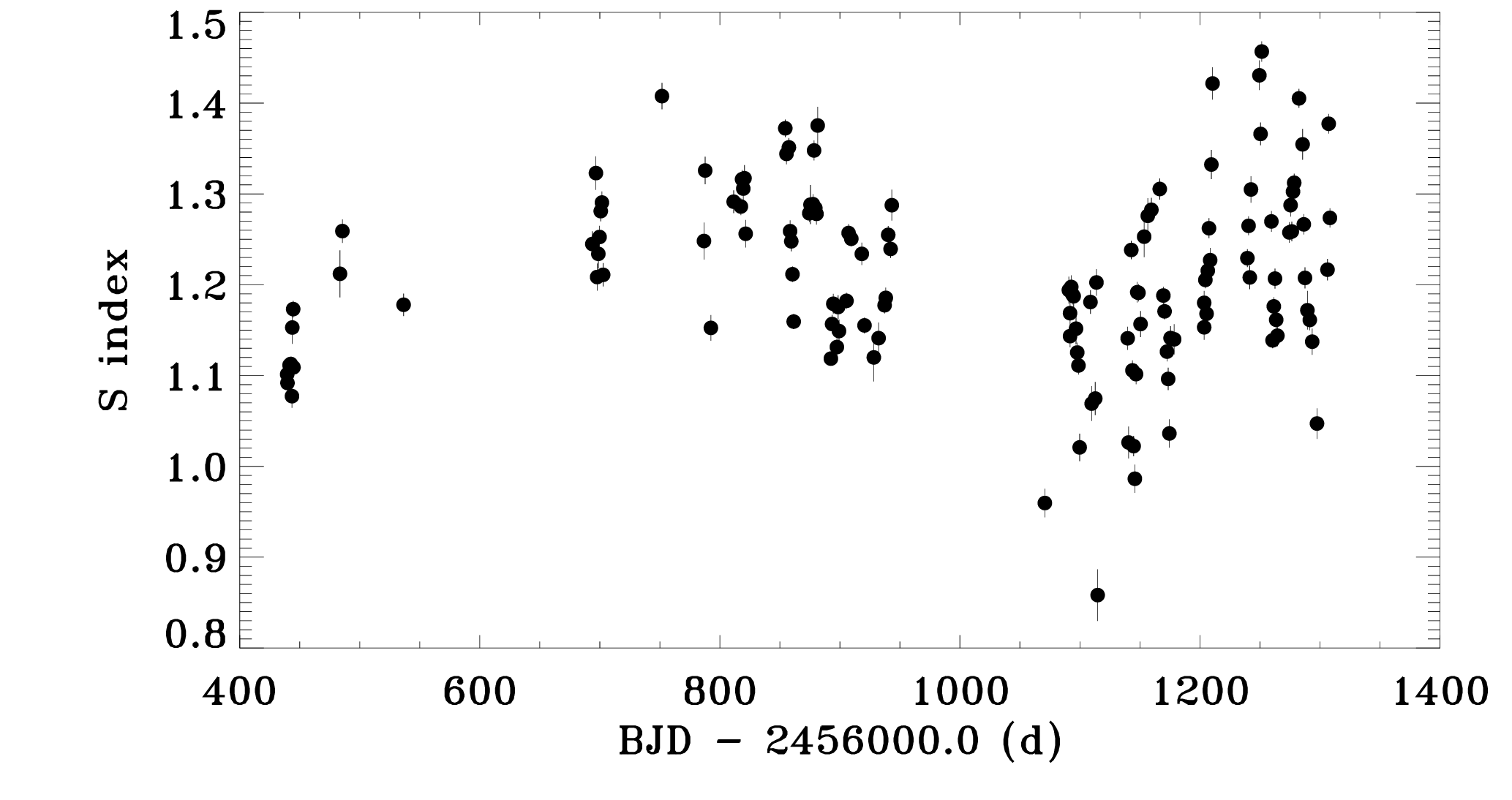}
\caption{Radial velocity (top) and activity indices H$\alpha$ (middle) and $S$ index (bottom) time series for GJ\,3998 measured with the TERRA
pipeline.}
\label{fig:rv}
\end{figure}

While the first planets discovered around M dwarfs were Jovian-type companions \citep{del98,mar98,mar01}, it is now rather well observationally established that their frequency is lower than that of giant planets around solar-type hosts \citep[e.g.][]{soz14}, and references therein), a result understood within the context of the core-accretion formation model \citep[e.g.][]{lau04}. The most recent evidence gathered by RV surveys and ground-based as well as space-borne transit search programs points instead towards the ubiquitousness of small (in both mass and radius) companions around M dwarfs. The outstanding photometric dataset of transit candidates around early M dwarfs from the Kepler mission has allowed \citet{gai16} and \citet{dre15} to derive cumulative occurrence rates of $2.3\pm0.3$ and $2.5\pm0.2$ planets with $1-4$ R$_\oplus$ for orbital periods $P$ shorter than 180 days and 200 days, respectively. Approximately one in two early M-type stars appears to host either an Earth-sized ($1-1.5$ R$_\oplus$) 
or a Super Earth ($1.5-2.0$ R$_\oplus$) planet \citep{dre15} with an orbital period $< 50$ days. Based on different recipes for the definition of the HZ boundaries and planetary atmospheric properties, \citet{kop13} and \citet{dre15} obtain frequency estimates of potentially habitable terrestrial planets ($1-2$ R$_\oplus$) around the Kepler early M-dwarf sample $\eta_\oplus=48^{+12}_{-24}$\% and $\eta_\oplus=43^{+14}_{-9}$\%, respectively. The inference from RV surveys of early- and mid-M dwarfs instead indicates $0.88^{+0.55}_{-0.19}$ Super Earth planets per star with $P< 100$ d and $m\sin i=1-10$ M$_\oplus$, and a frequency of such planets orbiting within the host's HZ $\eta_\oplus=41^{+54}_{-13}$\% \citep{bon13}. The $\eta_\oplus$ estimates from Doppler and transit surveys thus seem to be in broad agreement. However, $a)$ the uncertainties associated to these occurrence rate estimates are still rather large, and $b)$ the known compositional degeneracies in the mass-radius parameter space for Super Earths \citep[e.g.][]{rog10} make the mapping between the $\eta_\oplus$ estimates 
based on (minimum) mass and radius not quite straightforward. Any additional constraints coming from ongoing and upcoming planet detection experiments targeting M dwarfs are then particularly valuable. 

We present here high-precision, high-resolution spectroscopic measurements of the bright ($V=10.83$ mag) M1 dwarf GJ\,3998, gathered with the HARPS-N spectrograph \citep{cos12} on the Telescopio Nazionale Galileo (TNG) as part of an RV survey for low-mass planets around a sample of northern-hemisphere early-M dwarfs. The HADES (HArps-n red Dwarf Exoplanet Survey) observing programme is the result of a collaborative effort between the Italian Global Architecture of Planetary Systems \citep[GAPS,][]{cov13,des13,por16} Consortium, the Institut de Ciències de l'Espai de Catalunya (ICE), and the Instituto de Astrof\'isica de Canarias (IAC). Analysis of the Doppler time-series, spanning $\sim2.4$ years, reveals the presence of a systems of two Super Earths orbiting at 0.029 AU and 0.089 AU from the central star. After a short presentation of the HADES RV programme, in Sect.~\ref{2} we describe the observations and reduction process and in Sect.~\ref{obj} we summarize the atmospheric, physical and kinematic properties of GJ\,3998. In Sect.~\ref{3} we describe the RV analysis and discuss the impact of stellar activity. Sect.~\ref{sec:fot} presents the analysis and results of a multi-site photometric monitoring campaign. In Sect. \ref{sec:rvanalysis} we describe the Bayesian analysis, the model selection and derive the system parameters. We summarize our findings and conclude in Sect.~\ref{sec:concl}.

\section{The HADES RV programme}\label{2}

The HADES RV programme has completed the 6$^{th}$ semester of observations.
The complete original sample is composed by 106 stars, ranging from dM0 to dM3 spectral type, selected from the \citep{lep13} and Palomar/Michigan State University 
\citep[PMSU,][]{rei95} catalogs, with the additional criterion to be part of the APACHE catalog \citep{soz13}, with a visible magnitude lower than 12 and
with a high number of Gaia mission scans. The M stars are photometrically monitored also by the EXORAP program at Serra La Nave.
These selection criteria are meant to reach a good characterization
of the systems. From the complete sample, 27 stars were rejected during the first semester of observations (binary systems, fast rotators, peculiar stars, high
activity, earlier type stars and/or wrong spectral type).
Some of the targets ($\approx$\,15\% of the cleaned sample) have already more than 80 RV points, with an excellent precision (1-1.8 m~s$^{-1}$ at V$\approx$\,10-11). 
The star GJ\,3998 emerged among several possible candidates, since it shows clear periodic signals consistent with the presence of planets. \\
The statistical analysis of the capability of our survey is the purpose of the second paper of the HADES series (Perger et al. 2016, submitted). It is focused on simulating the planet detection rates on low-mass stars and predicting the number of planets likely to be detected, using the most recent planet occurrence statistics applied to our stellar sample and the actual observation times of our survey. The simulations performed by Perger et al., show, as first statistical outcomes, that: 1) the observations of the HADES program analyzed to date should enable the detection of $2.3\,\pm\, 1.6$\, planets (3.6\% detection rate), in agreement with our findings; 2) with 120 obs/star we are able to detect around 10\% of the distributed planets; 3) the optimum average number of observations per target, in case of time-limited survey, is around 35, for the case of early M-dwarf sample and exposure times of 900s (for further information see Perger et al. 2016, submitted, and references therein).

GJ\,3998 has been monitored from BJD = 2\,456\,439.6 (26th of May 2013) to BJD = 2\,457\,307.8 (12th of October 2015). We obtained a total of 136 data points spanning 869 days (Fig. \ref{fig:rv}).
The spectra were obtained at high resolution (R $\sim$\, 115 000) with the optical echelle spectrograph HARPS-N with exposure times of 15 minutes and average signal-to-noise ratio (S/N) of 45 at 5500 \AA. Of the 136 epochs, 76 were obtained within the GAPS time and 60 within the Spanish time. Observations were gathered without the simultaneous Th-Ar calibration, which is usually used to correct for instrumental drifts during the night. This choice avoided the contamination of the Ca II H \& K lines, which are particularly important for the stellar activity analysis of M dwarfs \citep{gia89,for09,lov11}. The M-type stars were observed by the Italian team in conjunction with other GAPS targets, which used the Th-Ar simultaneous calibration, thus we estimated the drift data between the two fibers (star and reference calibration) for each night from these observations, and evaluated the interpolated drift for GJ\,3998.

Data reduction and spectral extraction were performed using the Data Reduction Software \citep[DRS v3.7,][]{lov07}. RVs were measured by means of a weighted cross-correlation function (CCF) with the M2 binary mask provided with the DRS \citep{bar96,pep02}. This approach however is not optimal for M-dwarf stars, since their spectra suffer from heavy blends  which can lead to the mismatch of several features of the binary mask. As a result, the CCF shows sidelobes that affect the RV precision and the asymmetry indices of the CCF. A better alternative is to measure the RVs by matching the spectra with an high S/N template obtained by co-adding the spectra of the target, as implemented in the TERRA pipeline
\citep{ang12} which measures radial velocities by means of a match with a high S/N template and possibly provides a better RV accuracy when applied to M-dwarfs. 

%
\begin{table}[h]
\centering
\caption{Stellar parameters for the star GJ\,3998 from the analysis of the HARPS-N spectra using the technique in Maldonado et al.
(2015) (in the upper part). In the lower part of the table coordinates, V and K magnitudes, parallax, proper motions and space velocities are indicated. }             
\label{table:2}      
\begin{tabular}{ c c }     
\hline\hline      
Parameter$^{(4)}$ & GJ\,3998\\\hline
Spectral Type  & M1 \\  
$T_{\rm eff}$ [K]   & 3722$\pm$68   \\
$[{\rm Fe}/{\rm H}]$ [dex]    & -0.16$\pm$0.09  \\
Mass [M$_{\odot}$] & 0.50$\pm$0.05   \\
Radius [R$_{\odot}$] & 0.49$\pm$0.05   \\
log\,$g$ [cgs] & 4.77$\pm$0.04   \\
Luminosity [L$_*$/L$_{\odot}$] & 0.041$\pm$0.008   \\
v sin\,$i$ [km s$^{-1}$] & 0.93$\pm$0.55   \\
\hline
$\alpha$ (J2000) & $17^{h}$:16$^{m}$:$00.7^{s}$ \\
$\delta$ (J2000) & +$11^{o}$:$03'$:$30''$ \\
$V_{mag}$$^{(1)}$& 10.83 \\
$K_{mag}$$^{(2)}$& 6.82 \\
$\pi$[mas]$^{(3)}$& 56.20$\pm$2.26\\
$\mu_{\alpha}$[mas/yr]$^{(3)}$& -136.21$\pm$2.30\\
$\mu_{\beta}$[mas/yr]$^{(3)}$& -347.84$\pm$1.93\\
$U_{LSR}$ [km s$^{-1}$]$^{(4)}$ & -26.7$\pm$1.1 \\
$V_{LSR}$ [km s$^{-1}$]$^{(4)}$ & -52.8$\pm$1.3 \\   
$W_{LSR}$ [km s$^{-1}$]$^{(4)}$ & -28.8$\pm$0.6 \\    
S [km s$^{-1}$]$^{(4)}$ &  65.8$\pm$1.2\\\hline\hline                  
\end{tabular}
\begin{flushleft}
References. $^{(1)}$ \citet{koe10}; $^{(2)}$ \citet{cut03}; $^{(3)}$ \citet{vanl07}; $^{(4)}$ This work (see text).
\end{flushleft}
\end{table}
%


We list the data in Table~\ref{Tab1} including the observational dates 
(barycentric Julian date or BJD), the signal-to-noise ratios (S/Ns), the radial velocities (RVs) from the DRS and TERRA 
pipelines and the H$\alpha$ and $S$ indexes. 
We have an average S/N of 43, ranging from 18 
to 65 and a mean RV of -44.81 km s$^{-1}$. The TERRA RVs show a root mean squares (rms) dispersion of 4.24 m s$^{-1}$ and a mean error of  
1.12 m s$^{-1}$, whereas the DRS shows a dispersion of 4.89 m s$^{-1}$ and an error of 1.82 m s$^{-1}$. The standard deviation of the interpolated drift for GJ\,3998 is 0.7 m s$^{-1}$ which is less than the typical
DRS RV error of 1.8 m s$^{-1}$ and TERRA RV error of 1.1 m s$^{-1}$. The analysis described in the next sections has been performed on either the DRS and TERRA RVs and led to the same global results, with the latter providing better precision. In the following, only the results obtained with the TERRA RVs are listed. 

\section{Stellar properties of GJ\,3998}\label{obj}
GJ\,3998 is a high proper motion early-M dwarf (spectral type M1) at a distance of 17.8 pc from the Sun ($\pi$ = 56.20$\pm$2.26 mas). Accurate stellar parameters were determined using the empirical relations by \citet{mal15}\footnote{http://www.astropa.inaf.it/\textasciitilde{}jmaldonado/Msdlines.html} on the same spectra 
used in the present work to derive RVs. This technique relies on ratios of pseudo-equivalent widths of spectral features as a temperature diagnostic,
while combinations and ratios of features were used to derive calibrations for the stellar metallicity. The derived temperature and metallicity are used in association to
photometric estimates of mass, radius and surface gravity to calibrate empirical relationships for these parameters. From the Maldonado et al.'s study, typical uncertainties are in the order of 13.1\% for the stellar mass, 11.8\% for the radius, 25\% for luminosities, and 0.05 dex for log\,$g$. We note that these uncertainties were computed by taking into account the $\sigma$ of the corresponding calibration and the propagation of the errors in $T_{\rm eff}$ and [Fe/H]. \\ GJ\,3998 has a temperature $T_{\rm eff}$ = 3722 $\pm$ 68 K and a surface gravity log\,$g$ = 4.77 $\pm$ 0.04. \\
Stars presently near the Sun may come from a wide range of Galactic locations. Therefore, stellar space velocity, as a clue to the origin of a star in the Galaxy, is very
important. The accurate distance and proper motion available in the Hipparcos Catalogue (ESA 1997), combined with the stellar radial velocity, make it possible to derive reliable space velocities for GJ\,3998. The calculation of the space velocity with respect to the Sun is based on the procedure presented
by \citet{joh87}, corrected for the effect of differential galactic rotation \citep{sch88}, by
adopting a solar Galactocentric distance of 8.5 kpc and a circular velocity of 220 km s$^{-1}$. The correction of space velocity to the Local Standard of Rest is based on a
solar motion\footnote{In the present work, U is defined to be positive in the direction of the Galactic center.},
(U, V, W)$_{\odot}$ = (10.0, 5.2, 7.2) km s$^{-1}$, as derived from Hipparcos data by \citet{deh98}. The peculiar space velocity S, given by S = (U$^2$ + V$^2$ +
W$^2$)$^{1/2}$, is quoted with all kinematic
data and stellar properties in Table~\ref{table:2}. GJ\,3998 shows kinematic properties typical of the thin disk population. We have calculated the probabilities
that the star belongs to a specific population, thick (TD), thin disk (D) or stellar halo (H), following the method used by \citet{ben04}. 
On account of these probabilities GJ\,3998 could belong to the thin disk, because its ratio of the respective probabilities for the
thick and thin disks is TD/D $\lesssim$\, 0.5 \citep{ben04}.

\begin{figure}
\includegraphics[width=9cm]{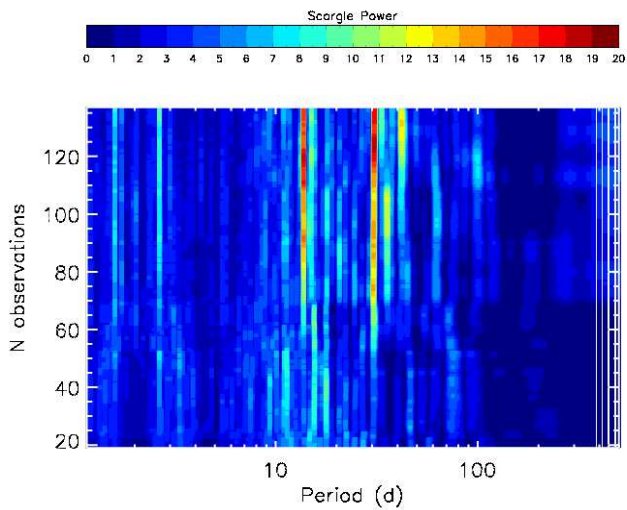}
\caption{Periodogram power of the inspected periods as a function of the number of observations. The most significant periods (the reddest ones) at 30.7 d and 13.7 d are clearly visible with a high power for N$_{obs}$ $>$\,60, as well as the period at 2.65 d with a moderate Scargle power, which increases for N$_{obs}$ $>$\,80 and the period at 42.5 d whose power increases for N$_{obs}$ $>$\,110  (2D plot adapted from A. Mortier - Private Communication).}
\label{fig:pow}
\end{figure}

\begin{figure}[!h]
\centering
\includegraphics[width=9.cm]{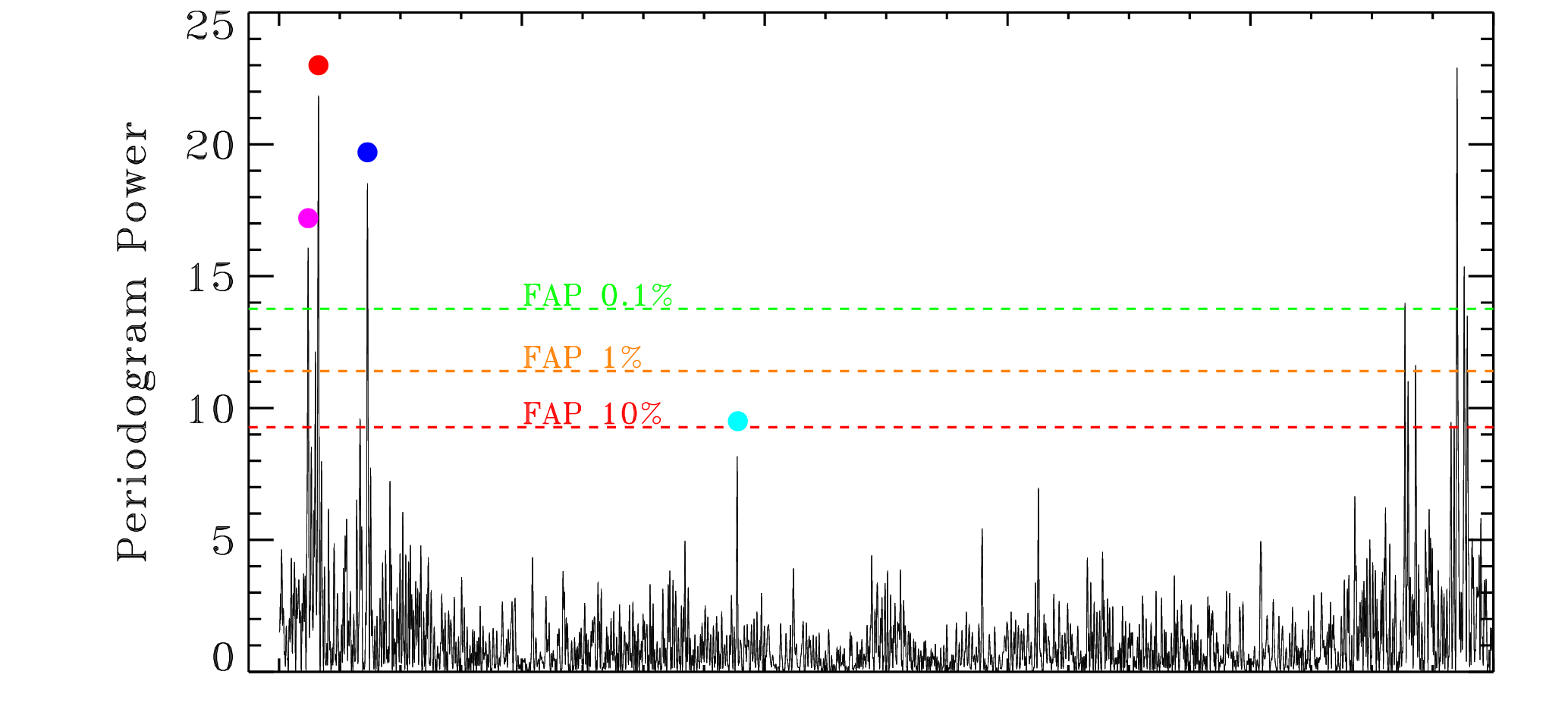}
\includegraphics[width=9.cm]{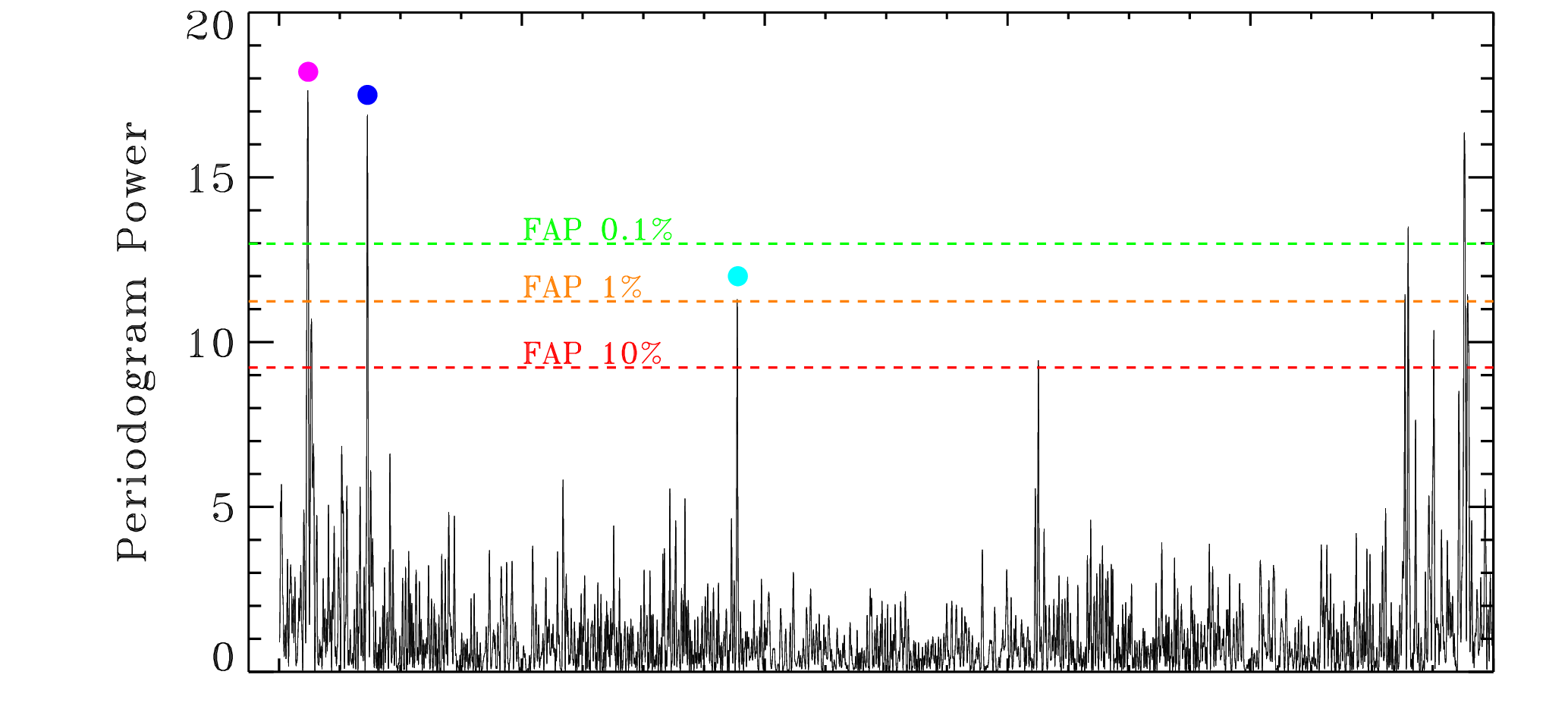}
\includegraphics[width=9.cm]{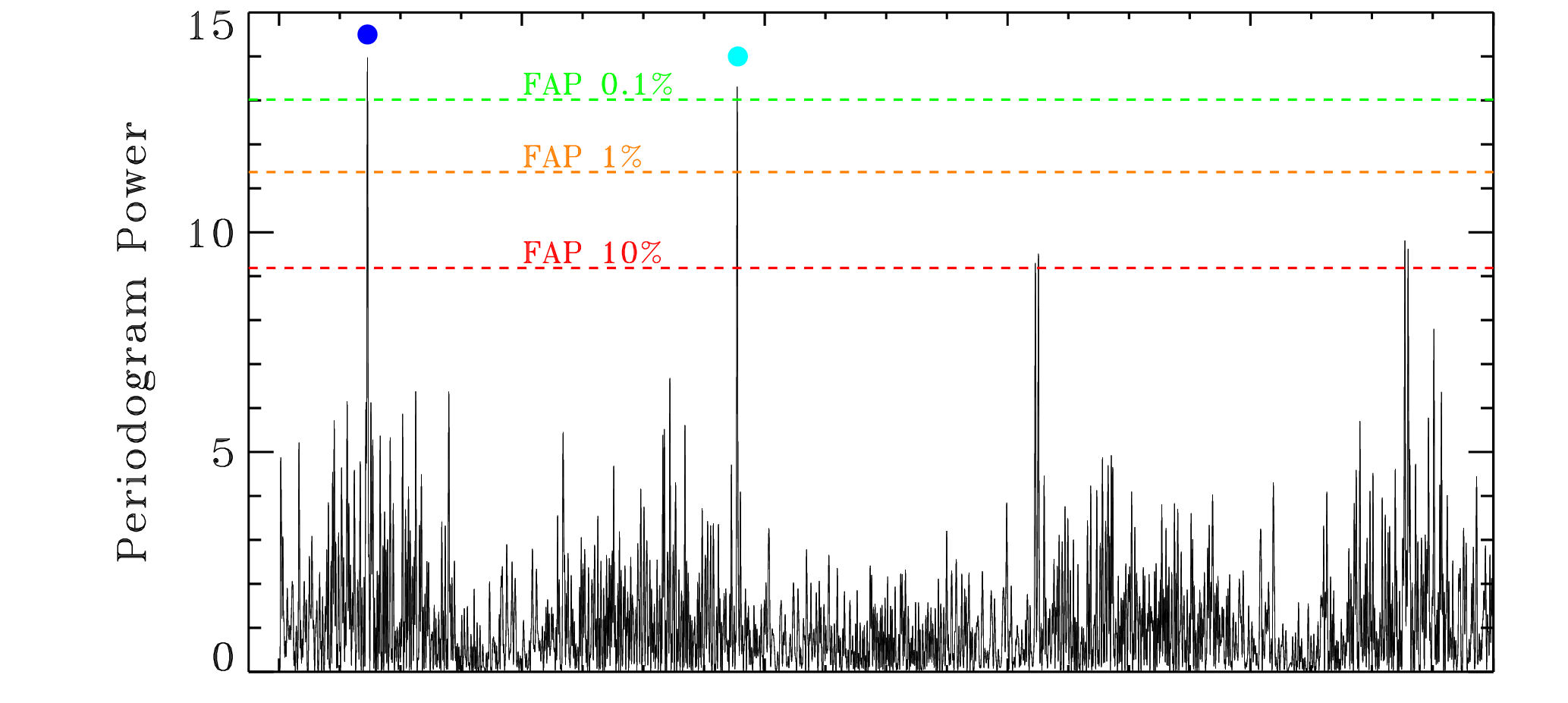}
\includegraphics[width=9.cm]{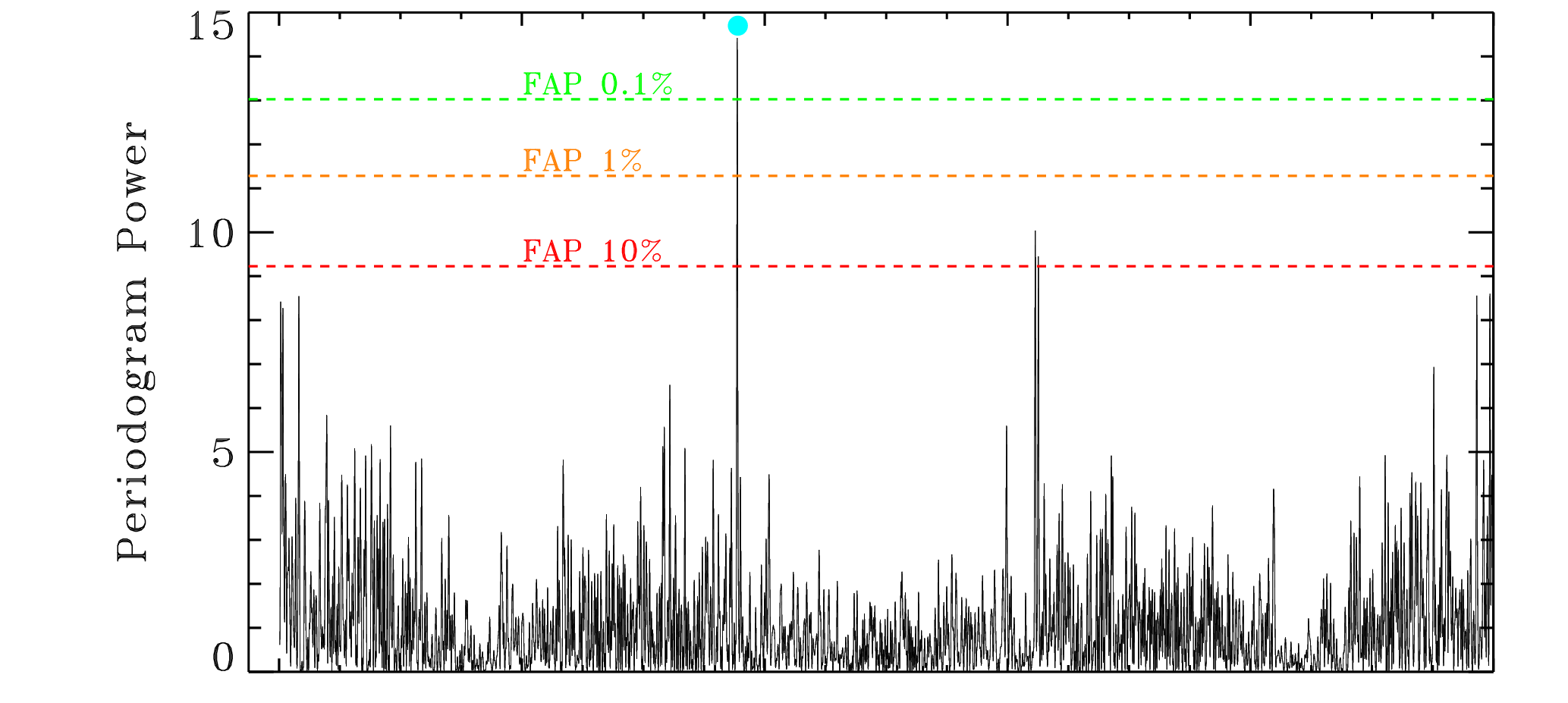}
\includegraphics[width=9.cm]{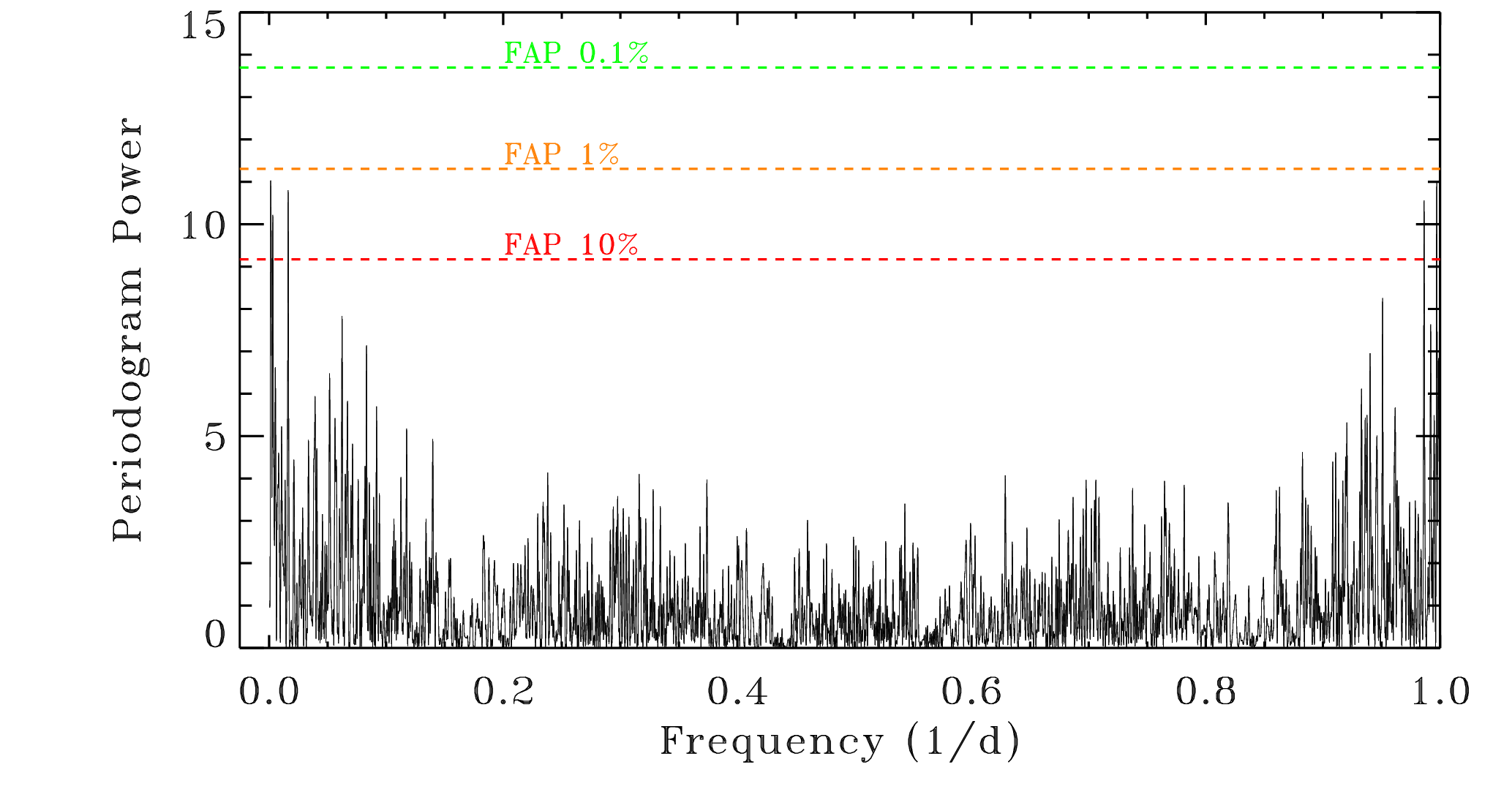}
\caption{GLS periodograms of the radial velocities of GJ\,3998, of the original data (upper panel) and after removing - from top to bottom - the 30.7 d (marked with the red dot), 
the 42.5 d (magenta dot), the 13.7 d (blue dot) and finally the 2.65 d (cyan dot) signals. The dashed lines indicate 0.1\%, 1\% and 10\% level of false alarm probability.}
\label{fig:per}
\end{figure}

\section{Data analysis}\label{3}

The raw rms dispersion of the
RVs is 4.24 m~s$^{-1}$ which is much higher than the average (Doppler + jitter) uncertainty <$\sigma_{i}$> $\approx$\, 1.4 -- 2.6 m~s$^{-1}$, depending on the
assumed jitter ($\sigma_{jitter}$ = 0.8 m~s$^{-1}$, from our first estimate from Gaussian Process, and $\sigma_{jitter}$ = 2.4 m~s$^{-1}$, following the derivation from Perger et al. 2016 (submitted), respectively). An F-test with F = $\sigma_{e}^2/<\sigma_{i}>^2$ \citep{zec09} returned a negligible probability ($< 10^{-8}$) that the photon noise combined with stellar jitter explains the measured dispersion.\\ 

The first step of the RV data analysis consists in identifying significant periodic signals in the data. Prewhitening is a commonly used tool for
finding multi-periodic signals in time series data. With this method we find sequentially the dominant Fourier components in a time series and remove them. The
prewhitening procedure was applied to the full RV data using the Generalized Lomb-Scargle (GLS) periodogram
algorithm \citep{zec09} and the program $\it Period04$ \citep{len04}, to have an independent test. The two methods give the same results in
terms of extracted frequencies from our RV time series. Due to the limited range in frequency covered by the found signals, we
repeated the analysis by using the iterative sine-wave fitting least-squares method \citep{van71}. This approach is used in the 
asteroseismic analysis of multiperiodic pulsating stars to minimize the subtle effects of prewhitening, such as power exchange between
a signal frequency and an alias of another signal frequency. After each detection, only the frequency values were introduced
as known constituents in the new search. In such a way their amplitudes and phases were recalculated for each new trial frequency,
always subtracting the exact amount of signal for any known constituent. This new analysis confirmed once more the frequencies 
previously found. \\ 
We calculated the false alarm probabilities (FAP) of detection using 10\,000 bootstrap randomization \citep{end01} of the original RV time series. 
Once a significant peak was located at a given period, the corresponding sinusoidal function was adjusted and removed. The process was repeated several times until no significant peak remained.\\
To visualize (Fig. \ref{fig:pow}) the cumulative contribution of data points to the significance of the detected frequencies, we calculated the GLS periodograms increasing the number of observations, from 20 to 136, adding one observation at a time (following the exact order of data acquisition). We compute the power corresponding to each of the 5000 periods in the range 1.2 d to 500 d, for each of the 117 periodograms obtained. Each horizontal slice of Fig. \ref{fig:pow} gives powers (indicated by the color scale) versus periods for the periodogram corresponding to the observations number, indicated on the vertical axis. Fig. \ref{fig:pow} is a 2D rapresentation adapted from A. Mortier (Private Communication).\\

Following this procedure, we identified three frequencies with FAP smaller than 0.1\%, at 0.0326 d$^{-1}$ (P\,=\,30.7 d), 0.0729 d$^{-1}$ (P\,=\,13.7 d) and 0.0235 d$^{-1}$ (P\,=\,42.5 d), in order of decreasing power. After removing
the sinusoid corresponding to the period of 30.7 d, with a semi-amplitude of 3.4 m~s$^{-1}$, two highly significant peaks are seen in the periodogram, at 42.5 d and
13.7 d, with semi-amplitudes of 2.5 and 2.1 m~s$^{-1}$ (Fig. \ref{fig:per}). One additional peak remains at 0.3773 d$^{-1}$ (P\,=\,2.65 d) with a
significant FAP smaller than  0.1\% and a semi-amplitude of 1.7 m~s$^{-1}$. A low frequency signal with a FAP  around 1\% remains after subtracting the four significant periods. This unresolved peak is probably the signature of a long-term variation of the activity, having a time scale of about 600~d or more. Indeed, this low frequency peak was found in both the activity indicators ($S$ index and H$\alpha$, Sect. \ref{sec:Prot}). Moreover, it almost disappears when introducing
a linear term in the frequency analysis of the RV time series.\\
A careful analysis of the spectral window, following the methods of \citet{daw10}, ruled out that the peaks in the periodograms are artifacts due to the combination of the long-term activity with the spectral window (or to the time sampling alone). \\

\subsection{Periodic signals: planet vs activity-related origin}\label{sec:Prot}

The purely frequentistic
approach highlighted the presence of four significant frequencies in the RV time series. However, when dealing with M-type stars we have to
face one of the major drawbacks of this kind of stars: many M stars show some level of activity, even if moderate, and display
inhomogeneities on their surface that rotate with the star. These inhomogeneities cause RV shifts due to the distortion of the
spectral line shape, and these shifts can $\it mimic$, confuse or even hide the planetary signal.\\

To investigate RV variability against stellar activity, we make use of spectral indices based on Ca II H\&K ($S$ index) and H$\alpha$ lines and we take
advantage of the photometric observations of the star within the APACHE and EXORAP photometric programs, to check whether plages or spots could produce the observed
Doppler changes (Sect. \ref{sec:fot}).\\ 

%
\begin{table}[h]
\centering
\caption{Ca II H \& K (top) and H$\alpha$ (bottom) windows.}             
\label{table:3}      
\begin{tabular}{ l l l }     
\hline\hline       
Band & Center (\AA) & Width (\AA) \\ 
\hline                    
Blue wing  & 3901.07 & 20 \\  
K core    & 3933.67 & 3.28 \\	
H core    & 3968.47 & 3.28\\
Red wing & 4001.07 & 20 \\\hline
Blue wing & 6550.87 & 10.75 \\
H$\alpha$\, core & 6562.81 & 1.6 \\
Red wing  & 6580.31  & 8.75 \\
\hline                  
\end{tabular}
\end{table}
%

Stellar activity is usually characterised by the strength of the central cores of the Ca II H \& K lines with respect to a reference flux measured at
each side of the lines. Activity $S$ indexes are measured by summing the fluxes in the central cores of the Ca II H \& K lines and rationing this sum to the sum of fluxes 
in two 20 \AA\, windows on either side of the lines. The definition of the windows is done following \citet{henry96} \citep{sua15}, and it is summarized in 
Table \ref{table:3}.
Fluxes were measured using the IRAF\footnote{IRAF is distributed by the National Optical Astronomy Observatories,
which are operated by the Association of Universities for Research
in Astronomy, Inc., under cooperative agreement with the National Science
Foundation.} task {\it sbands}. Before measuring the fluxes, each individual spectrum was corrected from its corresponding RV by using the IRAF task {\it dopcor}.
Uncertainties in the $S$ indexes were obtained by shifting the center of the red and blue wings by $\pm$ 0.2 \AA.
Typical uncertainties vary between 1$\cdot10^{-3}$ and 5$\cdot10^{-4}$ (Maldonado et al. 2016, submitted).\\

\begin{figure*}[h!]
\includegraphics[width=14cm]{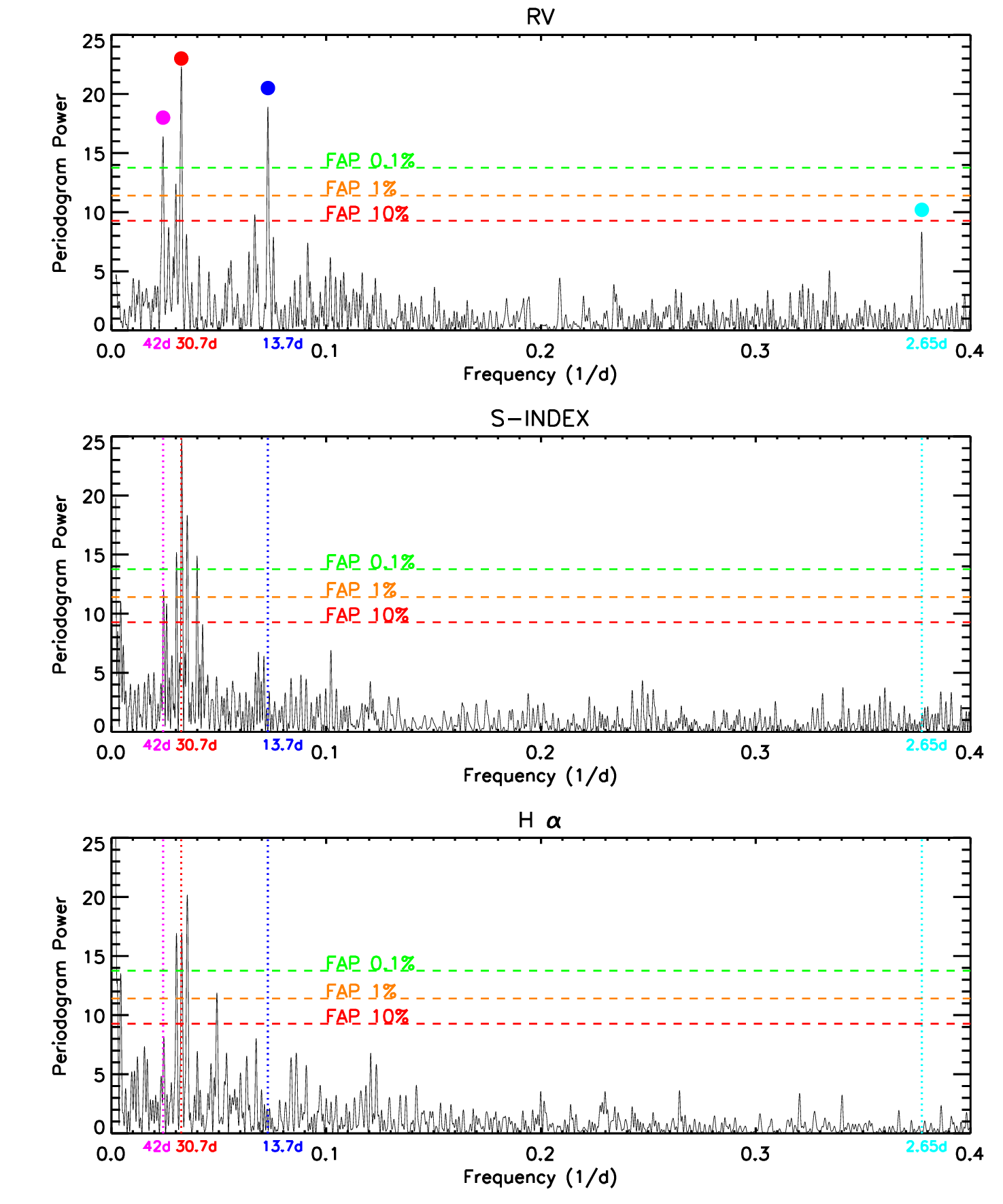}
\caption{From {\it Top} to {\it Bottom} the panels show the GLS periodograms of radial velocities, $S$ index and H$\alpha$, zoomed around the frequencies of interest. The dotted blue and cyan lines indicate the frequencies corresponding to orbital periods of the candidate planets at 13.7 d and 2.65 d, respectively, while the dotted red and magenta lines the frequencies corresponding to the activity periods at 30.7 d and 42.5 d, respectively. These frequencies/periods are marked with dots, with the same color code, in the upper RV panel.}
\label{fig:per_all}
\end{figure*}

H$\alpha$ emission is measured by summing the fluxes in the central core of the line at 6562.808\,\AA\, with a width of 1.6\,\AA\, and rationing this sum to the sum of fluxes in two windows on
either side of the line (see Table \ref{table:3}).
The H$\alpha$ line at $\lambda$\,= 6562.808\,\AA\, is sensitive to the mean chromospheric activity \citep{kur03,bon07,rob13}, as
Ca II H \& K lines.
The RV and activity indices H$\alpha$\, and $S$ index measurements of the star GJ\,3998 are displayed in Fig.\,\ref{fig:rv} as a function of time.\\
The analysis of the activity indicators was performed with the GLS algorithm and periodic variations around 30.7 and 42.5 d are indeed observed in the $S$ index and H$\alpha$, whereas no signal appears around 13.7 and 2.65 days, as illustrated in Fig. \ref{fig:per_all} showing the GLS periodograms of these parameters zoomed around the frequencies of interest. \\
The outcomes of the activity analysis support the
non-planetary origin of the 30.7 d and 42.5 d signals and give us an indication of the stellar rotation period and of a differential rotation, consistently with the outcomes of a recent analysis based on a wide dataset, including M stars, of $Kepler$ time series, as discussed in Sect. \ref{sec:concl}. This rotation period is consistent with the predicted Prot $\approx$\,$34.7\pm6.9\, d$ from the formula in \citet{sua15} according to the log$R'_{\mathrm{HK}}$ $\approx$\,-5.01.

\begin{figure*}[!h]
\includegraphics[width=15cm]{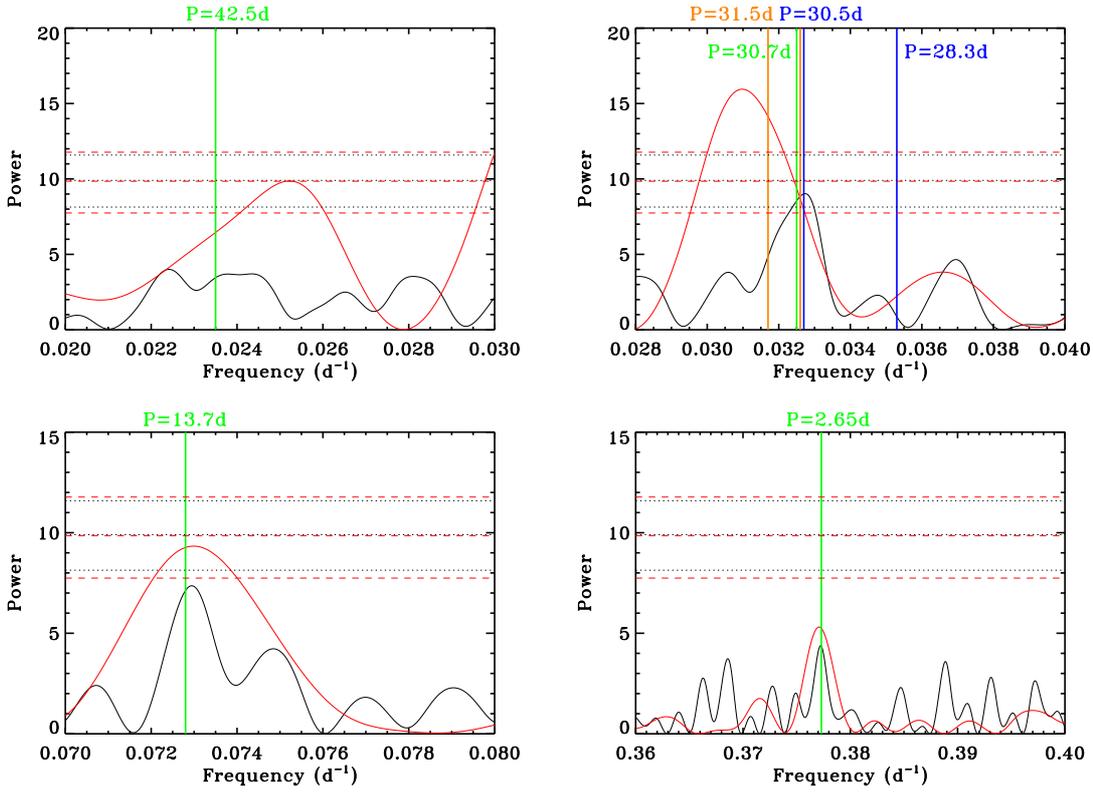}
\caption{GLS periodograms around the frequency ranges of interest, for two subsets of the original RV data. The first seasons has 61 data points and its periodogram is black (FAP: 0.1\%, 1\%, 10\% horizontal black dotted lines),
the second season has 75 data points and is plotted in red (FAP: 0.1\%, 1\%, 10\% horizontal red dashed lines). The two upper panels refer to the frequencies related to activity (f\,=\,0.0235 d$^{-1}$ -- P\,=\,42.5 d upper left panel; f\,=\,0.0326 d$^{-1}$ -- P\,=\,30.7 d upper right panel). The two lower panels refer to the frequencies related to the candidate planets (f\,=\,0.0729 d$^{-1}$ -- P\,=\,13.7 d lower left panel; f\,=\,0.3773 d$^{-1}$ -- P\,=\,2.65 d lower right panel). The relevant frequencies derived from all the methods used on the full dataset are indicated with vertical lines green from RV analysis, blue from activity indices and orange from photometry (see Sect. \ref{sec:fot}).}
\label{fig:sea}
\end{figure*}

In order to avoid any misinterpretation of the
stellar activity as a planetary signal, we analysed two subsets of the data (before and after BJD-2\,457\,000.0 d) to check the persistence of the planetary signals over time and to mitigate the possible effects
of discontinuities in the data sampling. In Fig. \ref{fig:sea} we show the GLS periodograms zoomed around the frequency ranges of interest, black-line for the first subset (61 RV
points) and red-line for the second subset (75 RV points). In the plot we indicated the frequencies obtained from the RV (green) and activity indices (blue) analysis and
from the photometry (orange) (see Sect. \ref{sec:fot}). Whilst the periods related to activity effects (related to magnetic phenomena, like spots and faculae, rotating on the stellar surface and the
presence of differential rotation) span a range from 28 to 41 d in the two seasons (f\,=\,0.036 d$^{-1}$ to f\,=\,0.024 d$^{-1}$), the two features at P$_{1}$ = 2.65 d (f\,=\,0.3773 d$^{-1}$) and P$_{2}$ = 13.7 d (f\,=\,0.0729 d$^{-1}$) are observed in both seasons. This constitute a further evidence that these two signals are present in the RVs at any time, leading us to the conclusion of a planetary origin for both signals.\\

\begin{figure}[!h]
\includegraphics[width=8.cm]{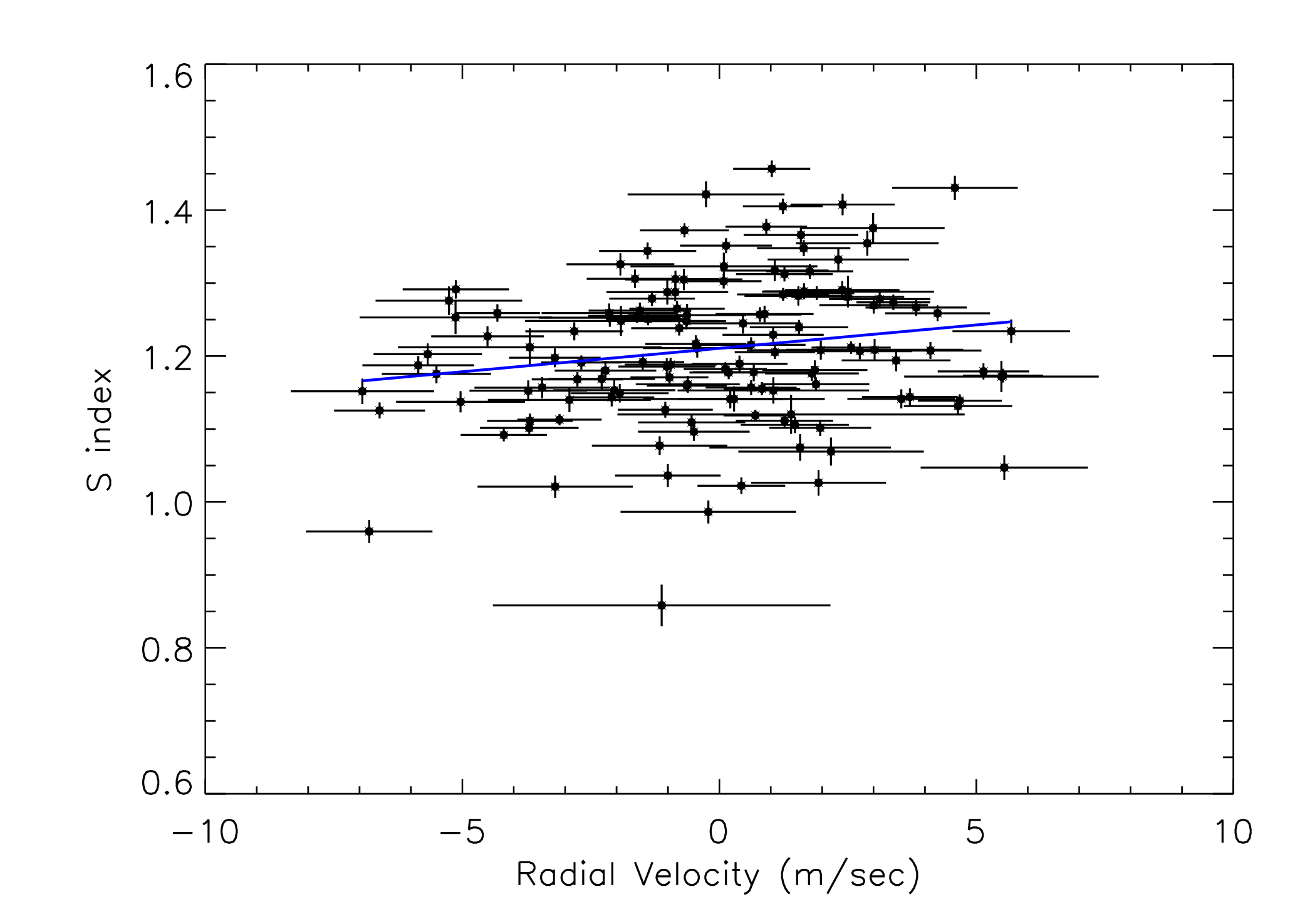}
\includegraphics[width=8.cm]{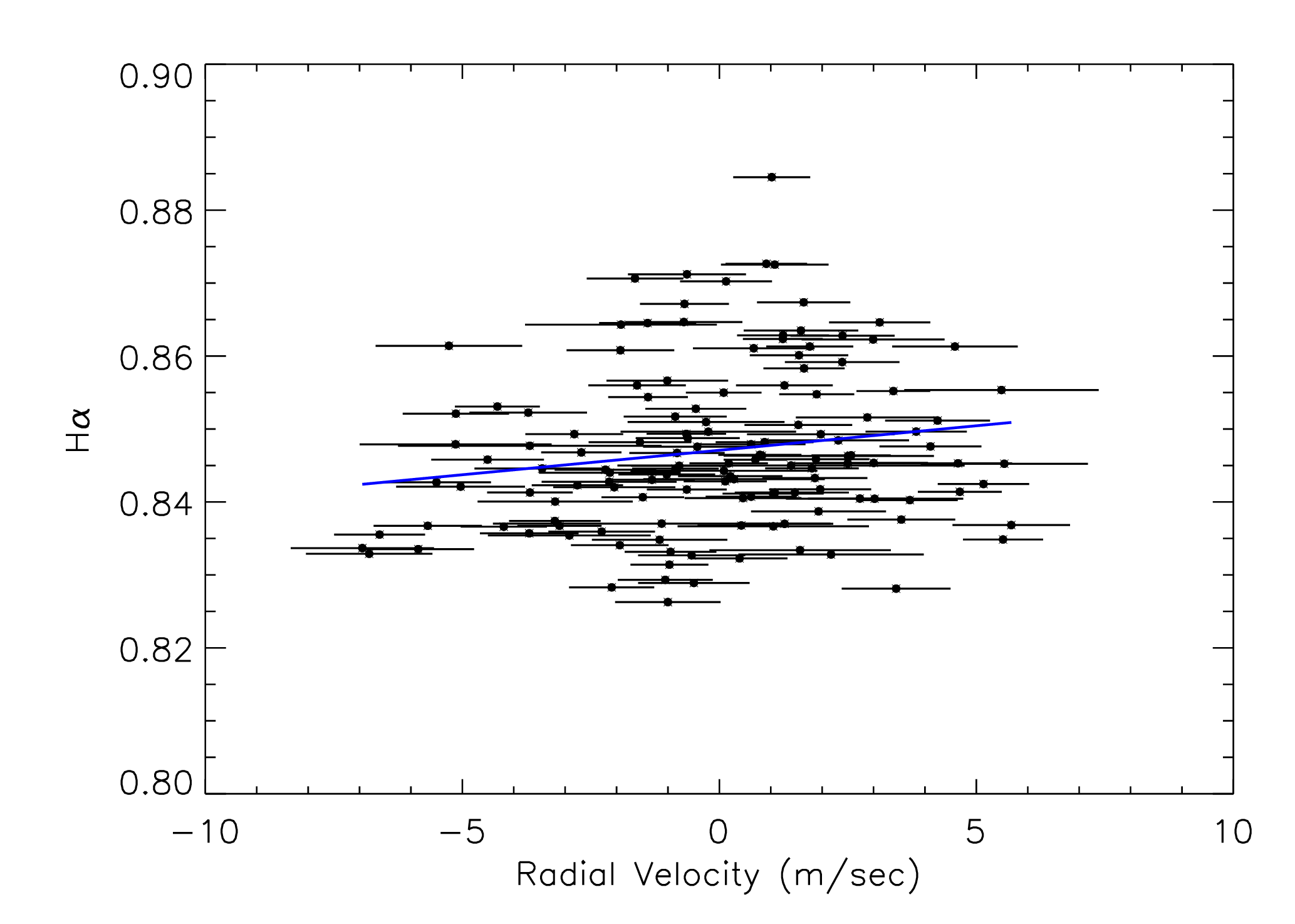}
\caption{Radial velocity after subtracting the contributions of the activity signals at 30.7 d and 42.5 d, displayed as a function of the $S$ index (top) and H$\alpha$
(bottom). The extremely low correlation values support the planetary interpretation of the 2.65 d and 13.7 d signals still present in the data.}
\label{fig:corr}
\end{figure}

Another argument in favour of the planetary interpretation of the 2.65 d and 13.7 d signals is provided in Fig. \ref{fig:corr} by the low Spearman's rank correlation coefficients between the S
index ($\rho$ = 0.17) and H$\alpha$ ($\rho$ = 0.18) and the RV residuals after removing the activity contributions (30.7 d and 42.5 d). Such a correlation would be expected if the radial velocity variation is still induced by activity features on the star surface.\\ 
The activity indices analysis has been performed with both the values calculated by the TERRA pipeline and those calculated, independently, using the method described above, obtaining the same results.\\

\section{Photometry}\label{sec:fot}

The targets of the HADES program are photometrically monitored by two independent programs: APACHE and EXORAP\footnote{EXOplanetary systems Robotic APT2 Photometry}.\\
These two programs regularly follow-up the sample of M stars to provide an estimate of the stellar rotation periods by detecting periodic modulation in the differential light curves.

\subsection{EXORAP photometry: dataset } 
\label{exorap}

GJ\,3998 was monitored at INAF-Catania 
Astrophysical Observatory with a 80cm f/8 Ritchey-Chretien robotic telescope (APT2)  located at Serra 
la Nave (+14.973$^{\circ}$E, +37.692$^{\circ}$N, 1725 m a.s.l.) on Mt. Etna. The APT2 camera is an 
ASPEN CG230 equipped with a 2k$\times$2k e2v CCD 230-42 detector operated with a binning factor of 2 
(pixel scale 0.94$"$) and a set of standard Johnson-Cousins $UBVRI$ filters. 
The data were reduced by  overscan, bias, dark subtraction and flat 
fielding  with the IRAF procedures, and visually inspected for  quality checking. We collected $BVRI$ photometry of GJ\,3998 in 
75 nights  between  5 May 2014 and Sept,17 2015. The log of the two observing seasons is given in Table~\ref{exorapobslog}.
We used aperture photometry as implemented in the IDL routine {\em aper.pro}, trying a range of apertures 
to minimize the rms of the ensemble stars.

To measure the differential photometry, we started with an
ensemble of $\sim$6 stars, the  nearest and brightest to GJ\,3998 and checked the variability of each of them building their differential light curves 
using the remaining ensemble stars as reference. With this method we selected the least variable stars of the sample, typically 4 of them.
The rms of the ensemble stars is $\lesssim 9$ mmag in all bands.

\begin{table}
\caption{EXORAP observations log for GJ\,3998}
\label{exorapobslog}
\begin{center}
\begin{tabular}{|c|c|c|c|c|c|c|}
\hline
Season & Start & End & \multicolumn{4}{|c|}{Data points}\\
 & & & B & V & R & I \\
 \hline
2014 & 5 May & 30 Jul & 26 & 25 & 24 & 32\\
2015 & 15 Jan & 17 Sep & 49 & 50 & 44 & 24\\
\hline
\multicolumn{3}{|c|}{All data} & 75 & 75 & 68 & 56\\
\hline
\end{tabular}
\end{center}
\end{table}

\subsubsection{EXORAP photometry: Searching for periodic variability}

The differential photometry of the target is analyzed using the GLS algorithm. We analyzed the full dataset and each season separately (Table~\ref{exorapobslog}). 
No data were rejected after evaluation for outliers (clip at 5$\sigma$). For each band, we analyzed the 
periodograms, assigning them the FAP using the bootstrap method with 10\,000 iterations. The periodograms of the $R$ and $I$ 
magnitudes do not show any peak above the $10\%$ significance level. 
The periodogram of the $B$ measurements is affected by a long-term trend: the combination with the spectral window aliasing
makes the determination of the periodicities, if any, very difficult. The results obtained from the $V$ measurements are more reliable.
We find a good confirmation of the 30.7~d period (P\,=\,$30.75\pm0.12\, d$, the precision is the formal error from the least square fitting), though flanked by the aliases at $\pm$1~cycle/year 
To check for possible systematic effects, we repeated the search for periodicity on all the measured stars in the field of view of GJ\,3998 (about 70 targets), but none of them shows significant variability with the same periodicity of our target star.

\begin{figure}
\centering
\includegraphics[width=8.5cm]{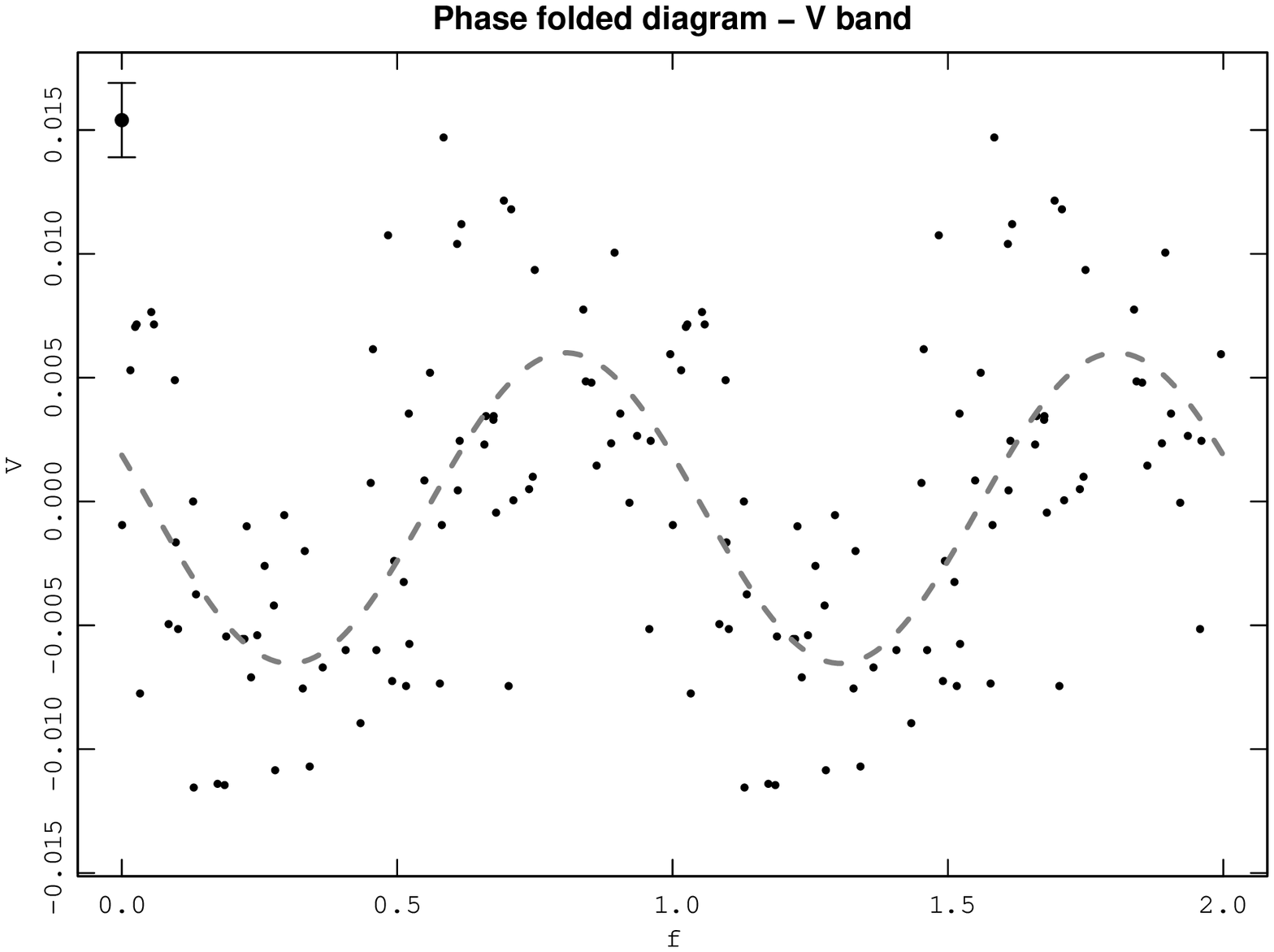}
\includegraphics[width=8.5cm]{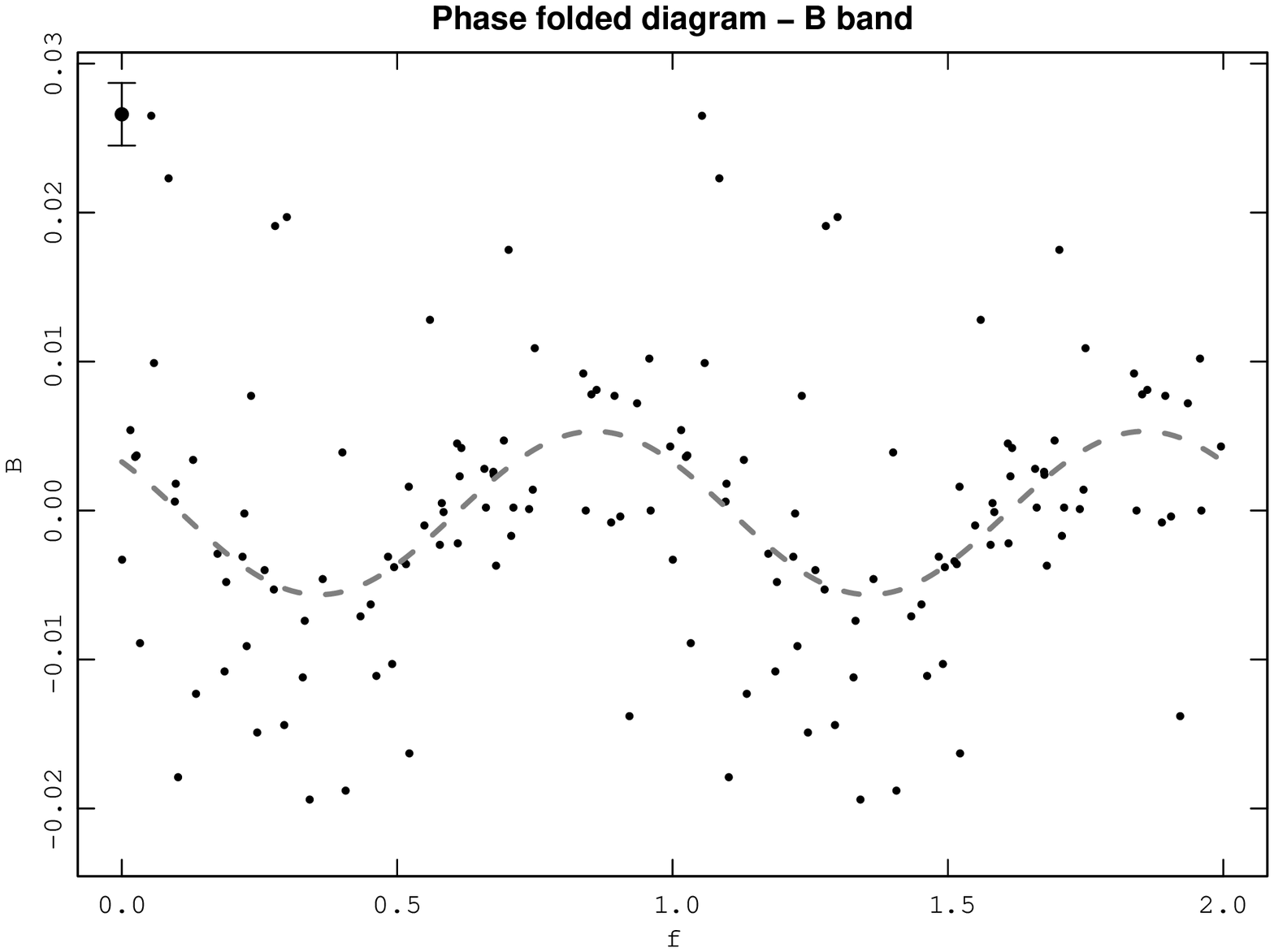}
\caption{Phase folded V and B curves at the periods P\,=\,$30.75\pm0.12\, d$, the precision reported is the formal error from the least square fitting. }
\label{lightcurves}       
\end{figure}

The  photometric period is consistent  with the one  found by the analysis of variability of features sensitive to
stellar activity in the HARPS-N data  (Sect. \ref{sec:Prot}).  The $V$ light curve folded with the observed periodicity is 
shown in Figure~\ref{lightcurves} (the $B$ light curve folded with the same periodicity is also shown). It shows a full amplitude of 0.012 mag.
We note that the observed variability amplitude in the $V$ band is typical for an M2 dwarf as shown by \citet{rock06}, 
who also report variability amplitudes in the $R$ and $I$ bands that are about half that of the $V$ band. With a precision of a few  millimag,  a signal with a five-millimag full-amplitude only remains elusive to us, hence the lack of period detection in the $R $ and $I$ data.

\subsection{APACHE photometry: dataset}\label{apachephotometry}
    APACHE is a photometric survey devised to detect transiting planets around hundreds of early-to-mid M dwarfs \citep{soz13}.
     GJ\,3998 was monitored for 41 nights between June 30, 2014, and August 13, 2015, with one of the five 40cm telescopes composing the APACHE array, located at the Astronomical Observatory of 
     the Autonomous Region of the Aosta Valley (OAVdA, +45.7895 N, +7.478 E, 1650 m.a.s.l.). Each telescope is a Carbon Truss f/8.4 Ritchey-Chretien equipped with a GM2000 10-MICRON mount 
     and a FLI Proline PL1001E-2 CCD Camera, with a pixel scale of 1.5 $^{\prime\prime}$/pixel and a field of view of 26$^{\prime}$x26$^{\prime}$. The observations were carried out using a 
     Johnson-Cousins \textit{V} filter following the standard strategy used by APACHE, consisting in three consecutive exposures repeated at intervals of $\sim$20-25 minutes, while the target 
     is $\sim35^{\circ}$ above the horizon. The images were reduced with the standard pipeline TEEPEE written in \texttt{IDL}\footnote{Registered trademark of Exelis Visual Information Solutions.} 
     by the APACHE team \citep[see][]{gia12}. TEEPEE is devised to perform ensemble differential aperture photometry by testing up to 12 different apertures and choosing the best set of 
     comparison stars that give the smallest rms for the differential light curve of the target.

\subsubsection{APACHE photometry: Analysis of the light curves}\label{lcanalysis}
\begin{figure}
   \centering
   \includegraphics[width=9cm]{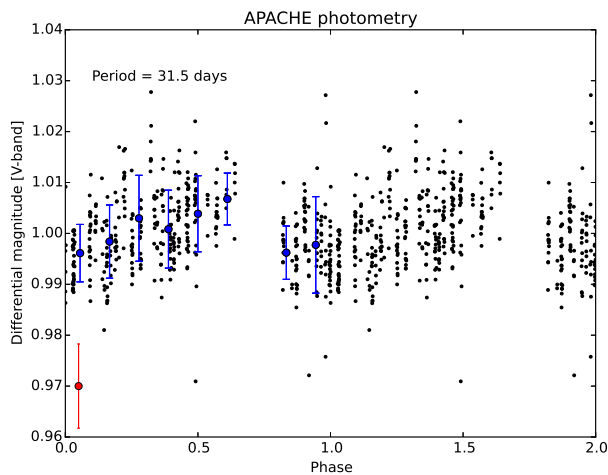}
   \caption{Differential light curve of GJ\,3998 observed by APACHE, folded at the best period P=31.5 days found through a frequency analysis of the time series. The red dot 
   in the lower left corner is to show the average uncertainty of the data.}
   \label{fig:apachefolded}%
    \end{figure}

The APACHE data were obtained by using an aperture of 4.5 pixel and a set of 4 comparison stars (UCAC4 506-065211, UCAC4 506-065134, 
 UCAC4 505-066479, and UCAC4 506-065155). The stellar brightness decreases almost linearly during the second season (Spearman's rank correlation coefficient $\rho=-0.66$, computed with the 
 \texttt{r$\_$correlate} IDL function), and the $\sim$6 mmag rms of the data, which is comparable to the typical photometric precision of the measurements, is not indicative of high photospheric 
 variability. The brightness decrease is not confirmed by the EXORAP $V$ photometry and hence its
origin remains unclear.
We searched the light curve for sinusoidal-like modulation by using the complete dataset consisting of 455 points, each being the average of three consecutive measurements with uncertainty equal 
 to their rms. We used the GLS and $\it Period04$ algorithms to calculate the frequency periodograms.
Gaps in the data and in the folded light curves make a little uncertain
the accurate identification of the true period around 30-35~d.
The analysis performed with $\it Period04$  returned a maximum frequency
corresponding to a period of about 31.5 d, in good agreement with the results of
the EXORAP photometry, while the GLS analysis returned the
1 cycle/year alias at 34.5 d. The best-fit semi-amplitude of the sinusoid is $\sim5$ mmag. Fig. \ref{fig:apachefolded} shows the light curve folded at the best period found. To estimate the significance of the 
 detection, we performed a bootstrap analysis (with replacement) by using 10\,000 fake datasets derived from the original photometric data, and found a FAP $\le$ 1.0$\%$. While indicating a rather 
 significant detection, this result should be taken with caution because the signal semi-amplitude is close to the typical precision of the data.
 However, it is noteworthy that the best 
 photometric frequency we find is close to that found in the RVs and spectroscopic activity index time series, which we explained as related to the stellar rotation frequency. 
 In sight of this, we conclude that the APACHE photometry further supports the interpretation based on spectroscopic evidence.    
 We searched the light curve also for possible transit-like signals of the candidate planets, but none was detected. However, this search is not conclusive since the number of observations and the photometric precision are too low to rule out signals in this period domain.

\subsection{FastCam Lucky Imaging Observations}

On October 1st 2014 at 22:04 UT, we collected 50,000 individual frames of GJ3998 in the \textit{I} band using the lucky imaging FastCam instrument \citep{osc08} at the 1.5m Carlos S\'anchez Telescope (TCS) at the Observatorio del Teide, Tenerife, with 30 ms exposure time for each frame. FastCam is an optical imager with a low noise EMCCD camera which allows to obtain speckle-featuring not saturated images at a fast frame rate, see \citet{lab11}.

\begin{figure} 
\centerline{\includegraphics[angle=0,scale=0.38]{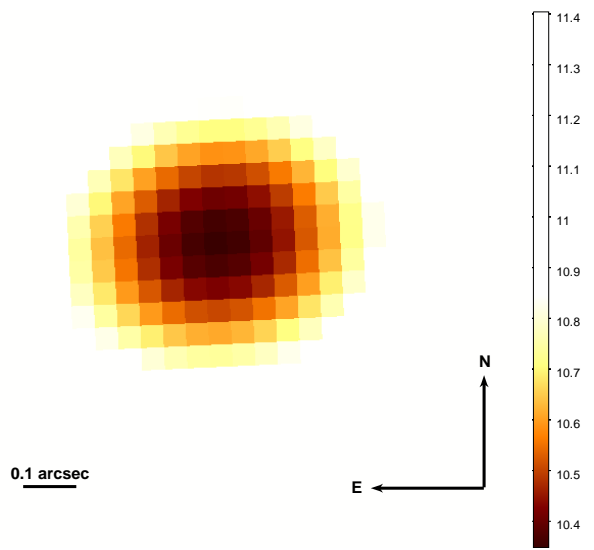}} 
\caption{Diffraction-limited image of GJ3998 after lucky imaging processing with a 20\% selection of the best individual TCS/FastCam frames.
\label{GJ3998}} 
\end{figure}

In order to construct a high resolution, diffraction limited, long-exposure image, the individual frames were bias subtracted, aligned and co-added using our own Lucky Imaging (LI) algorithm, see \citet{law06}. Fig. \ref{GJ3998} presents the high resolution image constructed by co-addition of the 20\% of the best frames, resulting in a total integration time of 300 seconds. The combined image achieved $\Delta m_I=2.5-3.0$ at $1\farcs0$, finding no bright contaminant star in the diffraction limited image.

\section{MCMC analysis of the radial velocity time series}
  	\label{sec:rvanalysis}

To derive the system parameters with the associated uncertainties and also to double check the purely frequentist approach, we performed a Bayesian analysis of the RV data by using the publicly available \texttt{emcee} tool
\citep{foreman13}, 
a Python implementation of the Affine Invariant Markov chain Monte Carlo (MCMC) Ensemble sampler. Under the hypothesis that up to four significant signals exist in the 
data (i.e. two of planetary origin and two ascribable to the stellar activity), as inferred from the analysis described in Sect. \ref{sec:Prot}, we tested several different models 
by changing the number of planets, assuming circular or eccentric orbits for the outer planet, and examining different ways to model the stellar activity noise. In all 
the cases we included an additional 'jitter' term summed in quadrature with the RV uncertainties. We used uniform priors for all the fitted parameters, by keeping 
unchanged their range of variability when passing from one model to the other. The best fit values are calculated as medians of the marginal posterior distributions, 
and the uncertainties represent their 16$\%$ and 84$\%$ quantiles. In modelling the RVs we ignored the mutual gravitational perturbations between the two planets.

\subsection{Modelling the stellar activity noise}
	\label{sec:rvstellaractivity}
The comparative analysis of the periodograms for the $S$ activity index and RVs, also supported by photometry, showed in the data a clear existence of a nearly periodic noise term of stellar origin, whose recurrence is likely ascribable to the stellar rotation. We take this evidence into account in our RV analysis by treating the short-term activity (i.e. related to the stellar rotation) in two different ways. 

As a first trial, we considered a strictly periodic term by fitting a sinusoid with period constrained between 25 and 45 days. 
Additionally, we used PyORBIT \citep{mal16} to test several combinations of sinusoids at the stellar rotational period and its harmonics, and independent amplitudes and phases for each observing season, without observing a significant improvement with respect to a single sinusoid approach.  

As a second, more sophisticated trial we treated the stellar activity term as correlated quasi-periodic noise, by using the approach based on Gaussian Processes (GPs; see, e.g. \citealt{rasmu06} and \citealt{roberts12}). In this case the stellar noise is conveniently described by means of a covariance function, opportunely selected, whose functional form and parameters have some correspondence to the physical phenomena to be modelled. The GP framework can thus be used to characterize the activity signal without requiring a detailed knowledge, for instance, of the distribution of the active regions on the stellar surface, their lifetime or their temperature contrast, which is almost always inaccessible.                   

To model the short-term activity in the GP framework we used the publicly available \texttt{GEORGE} Python library for GP regression \citep{ambi14}, and adopted the quasi-periodic kernel described by the covariance matrix

\begin{eqnarray} \label{eq:eqgpkernel}
k(t, t^{\prime}) = h^2\cdot\exp\large[-\frac{(t-t^{\prime})^2}{2\lambda^2} - \frac{sin^{2}(\pi(t-t^{\prime})/\theta)}{2w^2}\large] + \nonumber \\
+\, (\sigma^{2}_{RV}(t)\,+\,\sigma_{j}^{2})\cdot\delta_{t, t^{\prime}}
\end{eqnarray}
where $t$ and $t^{\prime}$ indicate two different epochs.
This kernel has been used with profit in several recent works on exoplanets to mitigate the impact of the stellar noise in the RV data \citep[see, e.g.,][]{hay14,grun15,faria16,mortier16}, because its functional form encompasses some properties of stellar activity that we can reasonably conceive. In fact, being composed by a periodic term coupled to an exponential decay term, this kernel describes a recurrent signal linked to the stellar rotation, and takes into account for a finite lifetime of the active regions and evolution of their size. The parameters of the covariance function (also called \textit{hyper} parameters) can be interpreted as follows:
\begin{itemize}
\item $h$ represents the amplitude of the correlations;
\item $\theta$ is usually related to the rotation period of the star (or one of its harmonics);
\item $w$ is the length scale of the periodic component, and can be linked to the size evolution of the active regions; 
\item $\lambda$ is the correlation decay timescale, and it can be related to the lifetime of the active regions.
\end{itemize}
Equation (\ref{eq:eqgpkernel}) also includes a term with the uncorrelated noise, added quadratically to the diagonal of the covariance matrix. Here,  $\sigma_{\rm RV}(t)$ is the RV uncertainty at time \textit{t}, $\sigma_{j}$ is the additional noise we fit in our models to take into account instrumental effects and other $\lesssim 1 \ms$ noise sources not included in the internal errors and in our stellar activity framework, and $\delta_{t, t^{\prime}}$ is the Dirac delta function.

Despite the activity index datasets suggest the existence of a long-term trend (i.e. hundreds of days), with properties that are far to be constrained by our data (see Sect. \ref{3}), the radial velocity time series shows a very weak similar trend, and for this reason we decided not to include it in our general RV model. 

In the GP framework, the log-likelihood function to be maximised by the MCMC procedure is 
\begin{equation} \label{eq:2-1}
\ln \mathcal{L} = -\frac{n}{2}\ln(2\pi) - \frac{1}{2}\ln(det\,\mathbf{K}) - \frac{1}{2}\overline{r}^{T}\cdot\mathbf{K}^{-1}\cdot\overline{r}
\end{equation}
where $n$ is the number of the data points, \textbf{K} is the covariance matrix built from the covariance function in Equation (\ref{eq:eqgpkernel}), and $\overline{r}$ represent the RV residuals, obtained by subtracting the offset and the Keplerian(s) signal(s) (i.e. the "deterministic" component) to the original RV dataset. 

A general form for the different models that we tested in this work is given by the equation
\begin{eqnarray}\label{eq:eqrvmodel}
\Delta RV(t) =  \gamma + \sum_{j=1}^{n_{planet}} \Delta RV_{Kep,j}(t) + \nonumber \\
                   +\, \Delta RV(t)_{\rm(activity,\, short-term)} \nonumber \\
				   = \gamma + \sum_{j=1}^{n_{planet}} K_{j}\large[\cos(\nu_{j}(t, e_{j}, T_{0j}, \textit{P}_{j}) + \omega_{j})  + e_{j}\cos(\omega_{j})\large]  + \nonumber \\
				  +\, GP\, \LARGE(\mathbf{or}\, A\cdot\sin\large[\dfrac{2\pi\,t}{P_{rot,*}} + \phi]\large\LARGE) 
\end{eqnarray}
where $n_{planet}=1,2$, $\gamma$ is the RV offset, $P_{\rm rot,*}$ is the stellar rotation period.
Instead of fitting the eccentricity $e_{j}$ and the argument of periapsis $\omega_{j}$ separately, we introduce the parameters 
$C_{j} = \sqrt{e_{j}}\cdot \cos \omega_{j}$ and $S_{j} = \sqrt{e_{j}} \cdot \sin \omega_{j}$ to reduce the covariance between $e_{j}$ and $\omega_{j}$ \citep{ford06}.

The choice in our analysis of the covariance function in Eq. (\ref{eq:eqgpkernel}) appears appropriate by looking at the results obtained from a GP analysis of the $S$ index time series. Here the "deterministic" component is the mean of the $S$ index. The best fit values of the hyper parameters are showed in Tab. \ref{table:Cahkparameters}. For $\theta$ we find a value which is in good agreement with the stellar rotation period previously estimated, and the value for $\lambda$ appears physically plausible for a Main Sequence star, because it indicates an evolutionary time scale of $\sim$1 month for the active regions, i.e. comparable to the stellar rotation period. Thus, the quasi-periodic kernel in Eq. (\ref{eq:eqgpkernel}) describes reliably the short-term stellar activity pattern imprinted in the $S$ index. 
 
For the MCMC analysis of the RVs, for each model we used 80 random walkers to explore the parameter space, with the GP hyperparameters $w$, $\lambda$, and $\theta$ randomly initialized from normal distributed values around the best estimates summarized in Tab. \ref{table:Cahkparameters}, using their uncertainties as $\sigma$ of the distributions. Same for the guess values of the planetary parameters $K$ and $P$, for which we used the estimates found through the preliminary periodogram analysis, while $T_{\rm 0,conj}$ was initialized to an intermediate value in the prior range we defined. We applied an initial burn-in by definitively removing the first 3,000 steps from each chain. After the selection of the best model, based upon the estimates of the Bayesian evidences (see below), we performed an additional burn-in to derive the parameter estimates by removing from the full list of the posterior samples those having a $\ln \mathcal{L}$ lower than the median of the whole $\ln \mathcal{L}$ dataset. Typically, the convergence of the MCMC chains has been reached after $\sim$35,000 steps of the random walkers, with parameter correlation lengths $\sim$1,000-2,000.

   \begin{figure}
   \centering
   \includegraphics[width=9cm]{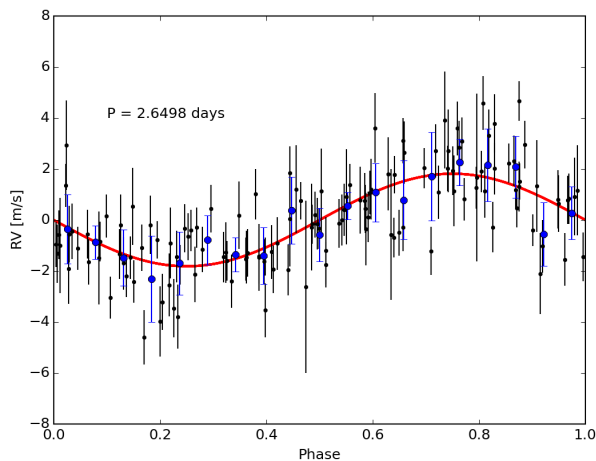}
   \caption{RV data folded at the best-fit orbital period of the inner planet. The signal of the outer planet, the RV offset and the best-fit GP solution for the stellar activity noise have been subtracted. Blue dots are the mean values in bins of amplitude 0.05. The red solid line is the best-fit orbital solution.}
   \label{fig:innerfase}%
    \end{figure}
%

   \begin{figure}
   \centering
   \includegraphics[width=9cm]{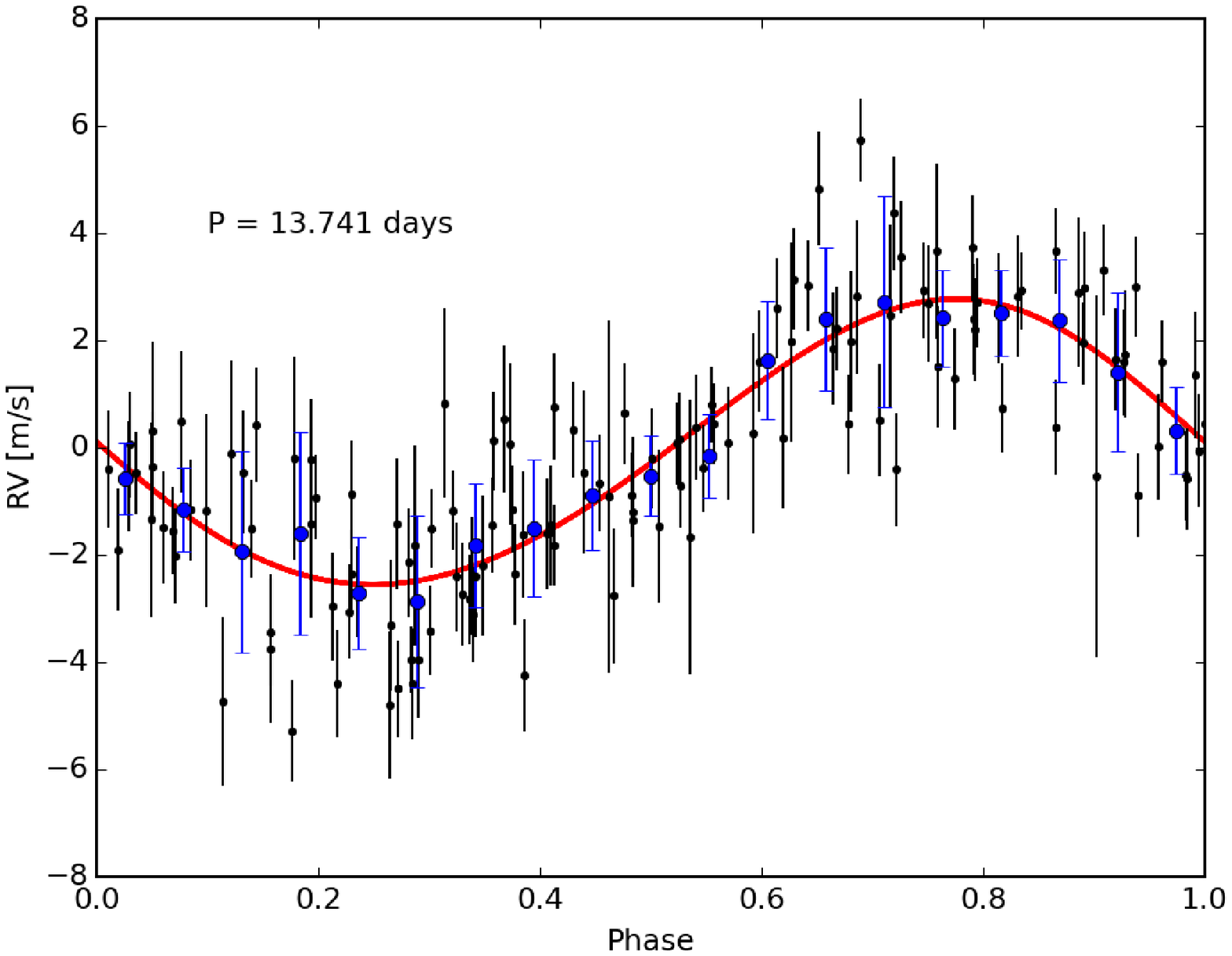}
      \caption{RV data folded at the best-fit orbital period of the outer planet. The signal of the inner planet, the RV offset and the best-fit GP solution for the stellar activity noise have been subtracted. Blue dots are the mean values in bins of amplitude 0.05. The red solid line is the best-fit orbital solution.}
   \label{fig:outerfase}%
    \end{figure}
%

   \begin{figure*}
   \centering
   \includegraphics[width=15cm]{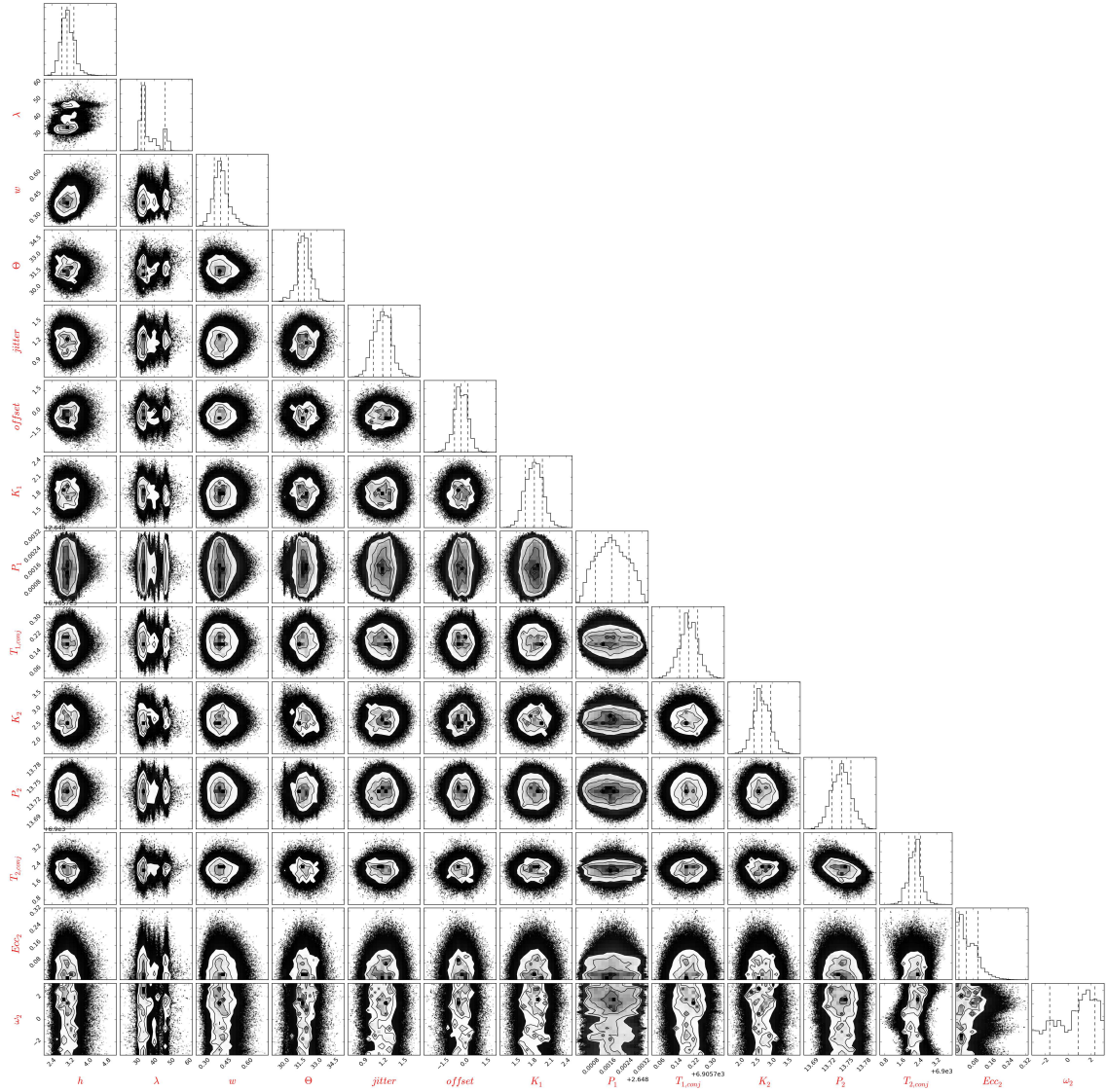}
      \caption{Marginal posterior distributions for the parameters of our model, which include two Keplerians and a GP-based fit for the short-term stellar noise. Vertical dashed lines indicate the median values and the 16$\%$ and 84$\%$ quantiles of the distributions.}
   \label{fig:posteriors}%
    \end{figure*}
%

   \begin{figure}
   \centering
   \includegraphics[width=9cm]{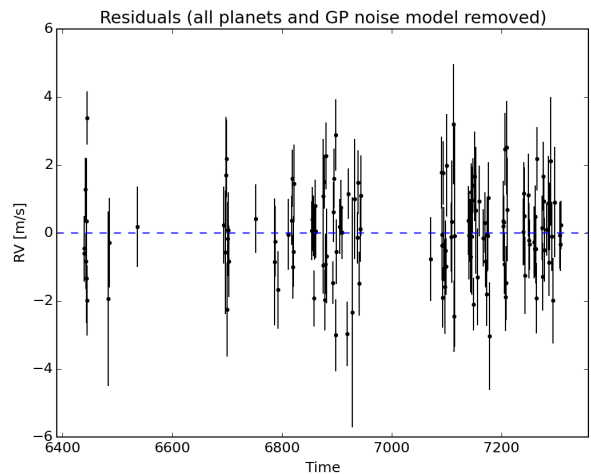}
   \caption{Residuals of the global fit (activity+Keplerians) we used to model the RV data set.}
   \label{fig:rvres}%
    \end{figure}

\subsection{Model selection and planetary parameters}
	\label{sec:rvsmodel}

Following the principles of Bayesian inference, for each of the tested models we calculated the logarithm of the Bayesian evidence $\mathcal{Z}$ to perform a model comparison and selection. To this purpose, because the estimate of $\mathcal{Z}$ is notoriously a challenging numerical task, we used two different estimators proposed in literature (\citealt{chib01}, hereafter C$\&$J estimator; and \citealt{perra14}, hereafter Perr estimator). For the interpretation of the model probabilities we follow here a conventional empirical scale often used (a slightly modified version of the so-called "Jeffreys' scale"), which states that the model with the highest $\mathcal{Z}$ is strongly favoured over the other when $\Delta\ln\mathcal{Z}\geqslant5$, while $2.5<\Delta\ln\mathcal{Z}<5$ denotes moderate evidence, $1<\Delta\ln\mathcal{Z}<2.5$ weak evidence, and $\Delta\ln\mathcal{Z}<1$ corresponds to inconclusive evidence. 

We report in Tab. \ref{table:modelcomparison} only the Bayesian evidences $\ln\mathcal{Z}$ for the models involving two Keplerian signals, which will be the object of our discussion. In fact, as expected from our previous analyses, models involving only one planet appear much less probable than those with two planets, independently on the framework used to account for the stellar activity. For instance, for circular orbits and considering only the C$\&$J estimator, $\Delta\ln\mathcal{Z}\sim17$ between the 2-planet model and that with only the outer companion, $\Delta\ln\mathcal{Z}\sim5$ when the model with only the inner planet is considered\footnote{We note that the significant difference between the evidences $\ln\mathcal{Z}$ of the 1-planet models is reflected in what is observed in the model residuals. In fact, when analysing with GLS the residuals of the 13.7-day planet model, the highest peak in the periodogram appears at the orbital period of the inner planet, and the estimate of signal semi-amplitude is $\sim1.6 \ms$, very close to our adopted value for the Keplerian semi-amplitude, meaning that the model can be improved by including an additional signal. On the contrary, the residuals of the 2.6-day planet model still contain a signal at the orbital period of the outer planet, but this is not the most significant and has a semi-amplitude of just $\sim0.3 \ms$, indicating that the GP model has significantly absorbed it, nonetheless without removing it. Analysing the results of the MCMC for the 2.6-day planet model, we notice that the hyperparameter theta is ~29+/-0.5 days, a value which is very close to double the orbital period of the outer planet. Then, it is likely that the rotational term of the GP is mostly responsible for the suppression of the 13.7 day signal.}. 
   
In Tab. \ref{table:rvparameters} we summarize the results for the final model we selected. The short-term stellar activity is modelled better in the GP framework with respect to a sinusoid. For instance, for 2-planet circular models the C$\&$J estimator results in $\Delta\ln\mathcal{Z}=\ln\mathcal{Z}_{\rm GP}-\ln\mathcal{Z}_{\rm sin}\sim16$. The reason probably lies in the lifetime of the active regions $\lambda$, which is too short to be properly modelled by a strictly periodic function.

We adopt here the model with the eccentricity of the outer planet treated as a free parameter (model number 3 in Tab. \ref{table:modelcomparison}). According to the statistical evidence, it is as likely as the simpler model with circular orbits (number 2), but we select it for convenience despite it involves a greater number of free parameters. In fact, by adopting model 3 we provide an upper limit on the outer planet eccentricity that can be useful for statistical studies, despite our estimate of $e$ is indicative of a circular orbit. 
For completeness, we also tested the model where the eccentricities of both planets are treated as free parameters (model 4 in  Tab. \ref{table:modelcomparison}). Based on a simple tidal dissipation model, we expect that the orbital eccentricity of the inner planet should be negligible. In fact, if the modified tidal quality factor of the inner likely super-Earth planet $Q^{\prime} \sim 1400 $, i.e., similar to that of the Earth in the case of the semidiurnal tide raised by the Moon \citep{lainey16}, and the outer planet does not significantly excite the eccentricity of its orbit, the tidal dissipation inside the inner planet can circularize its orbit with an e-folding timescale of only $\sim 65$ Myr, i.e., much shorter than the likely age of the system, as we can reasonably deduce from the quite long stellar rotation period. Our analysis results in $e=0.36^{+0.10}_{-0.12}$ for the eccentricity of planet $b$, and $e=0.061^{+0.059}_{-0.039}$ for that of planet $c$ (16$\%$ and 84$\%$ quantiles), with a Bayesian evidence which is slightly worse or slightly better than model 3, depending on the estimator (Tab. \ref{table:modelcomparison}). Thus, there is at most only a weak statistical evidence to prefer model 4 over model 3, with the latter that has fewer free parameters and therefore it is adopted also because simpler.
In Figs. \ref{fig:innerfase} and \ref{fig:outerfase} we show the RV data folded at the best-fit orbital period of the inner and outer planets, respectively, as well as the best-fit orbital solution. In each plot the signal of the other planet, the RV offset and the best-fit GP solution for the stellar activity noise have been subtracted. In Fig. \ref{fig:posteriors} we show the marginal posterior distributions for the parameters of the selected model. The RV residuals are showed in Fig. \ref{fig:rvres}. No residual long-term trend is visible in the data, supporting our a-priori decision of not including an additional term to model a possible stellar activity cycle. The analysis of the residuals with the GLS algorithm does not show any significant periodicity left in the data, meaning that the GP has absorbed the signal at $\sim42$ days. This could be related to the best-fit value assumed by the $\lambda$ time scale. In fact, the upper uncertainty on $\lambda$ is quite high ($\sim11$ days) and this is related to the presence of a secondary peak visible in the marginal posterior distribution for $\lambda$ slightly higher than 40 days. 

When looking at the best-fit values of the covariance function hyper parameters, that are showed in Tab. \ref{table:rvparameters}, we note that the distributions of $\lambda$ and $\theta$, which was constrained within a range that includes the two 'activity' periods found through the periodogram analysis, appear quite similar to those for the $S$ index. Because of their physical interpretation, one can expect them to remain almost unchanged when passing from the $S$ index to the RVs \textit{if the structure of the correlated (stellar) noise is similar in both data sets}. Moreover, both for the $S$ index and RVs $\theta$ finely settles to a value which is compatible with the estimates of the stellar rotation period previously derived. Thus, there is a convincing evidence that the use of a GP efficiently mitigate the short-term activity noise present in the RV data, resulting in reliable estimates for the orbital and physical parameters of the planets.


    \begin{table}
    \caption{Best fit values for the hyper parameters of the covariance function in Eq. \ref{eq:eqgpkernel}, derived from a GP analysis of the $S$ activity index time series using 100 random walkers. They are calculated as medians of the marginal posterior distributions, and the uncertainties represent their 16$\%$ and 84$\%$ quantiles. The priors used in our MCMC fitting are also showed.}
    \label{table:Cahkparameters}
	\centering
	\begin{tabular}{lll}
	\hline
    \noalign{\smallskip}
    Hyper parameter  &  Value &  Prior \\
    \noalign{\smallskip}
    \hline
    \noalign{\smallskip}
	$\lambda$ [days] &   $30.9\pm4.9$  & \textit{$\mathcal{U}$}(0,+$\infty$) \\ [3pt]
	$w$     &  $0.099\pm0.011$  &  \textit{$\mathcal{U}$}(0,+$\infty$) \\ [3pt]
	$\theta$ [days] &  $29.2\pm0.19$  &  \textit{$\mathcal{U}$}(25,50)\\ \noalign{\smallskip}
	 \hline
	\end{tabular}
    \tablefoot{The symbol \textit{$\mathcal{U}$}($\cdot,\cdot$) denotes an uninformative prior with corresponding lower and upper limits.}
    \end{table}
%
 

     \begin{table*}
    \caption{Summary of the Bayesian statistical evidences $\ln\mathcal{Z}$ for the two-planet models tested in this work. References to the three different methods used are given in the text. }
    \label{table:modelcomparison}
	\centering	
	\begin{small}
	\begin{tabular}{llllll}
	\hline
    \noalign{\smallskip}
    Model nr.  &  Nr. of free param. &  Eccentric/circular orbit & Stellar activity model & $\ln\mathcal{Z}_{C\&J}$ & $\ln\mathcal{Z}_{Perr}$ \\
    \noalign{\smallskip}
    \hline
    \noalign{\smallskip}
     1 & 11 & c+c & Sinusoid + jitter & -332.5 & -340.4 \\ [3pt]
	 2 & 12 & c+c & GP + jitter & -316.8 & -316.7 \\ [3pt]
	 3 & 14 & c+e & GP + jitter & -314.2 & -315.5 \\ [3pt]
	 4 & 16 & e+e & GP + jitter & -315.4 & -313.6 \\\noalign{\smallskip}
	 \hline
	\end{tabular}
	\end{small}
    \end{table*}
%


    \begin{table*}
    \caption{Best-fit values for the parameters of the global model (two planets\,+\,stellar activity) selected as explained in \ref{sec:rvsmodel}. They are calculated as medians of the marginal posterior distributions, and the uncertainties represent their 16$\%$ and 84$\%$ quantiles, unless otherwise indicated. The priors used in our MCMC fitting are also showed.}
    \label{table:rvparameters}
	\centering
	\begin{tabular}{lll}
	\hline
    \noalign{\smallskip}
    Parameter   &  Value &  Prior \\
    (planets) & & \\
    \noalign{\smallskip}
    \hline
    \noalign{\smallskip}
    \textbf{Planet $b$} & & \\ [3pt]
	$K$ [$\ms$] &   $1.82^{+0.14}_{-0.16}$  & \textit{$\mathcal{U}$}(0,$+\infty$) \\ [3pt]
	$P$  [days]   &  $2.64977^{+0.00081}_{-0.00077}$  &  \textit{$\mathcal{U}$}(1.5, 3.5) \\ [3pt]
	$T_{\rm 0,conj.}$ [BJD-$2,450,000$] &  $6905.895^{+0.042}_{-0.040}$  &  \textit{$\mathcal{U}$}(6900, 6915)\\ [3pt]
	$e$ & 0 (fixed) & - \\ [3pt]
	$a$ [au] & $0.029\pm0.001$ & derived\\ [3pt]
	$m_{\rm p}\sini$ [$\mearth$] & $2.47\pm0.27$ & derived \\[10pt]
	\textbf{Planet $c$} & & \\ [3pt]
	$K$ [$\ms$] &   $2.67^{+0.28}_{-0.23}$  & \textit{$\mathcal{U}$}(0,$+\infty$) \\ [3pt]
	$P$  [days]   &  $13.740\pm0.016$  &  \textit{$\mathcal{U}$}(12, 15.5)\\ [3pt]
	$T_{\rm 0,conj.}$  [BJD-$2,450,000$] &  $6902.2^{+0.24}_{-0.29}$  &  \textit{$\mathcal{U}$}(6900, 6915)\\ [3pt]
	$e$ & $0.049^{+0.052}_{-0.034} $ &  \\ [3pt] 
	 & $0.049^{+0.094}_{-0.045}$ ($5\%$\, and\, $95\%$ quantiles) & \\ [3pt]
	$\omega$ [rad] & unconstrained & \\ [3pt]
	$\sqrt{e}\cdot\sin\omega$ & $0.079^{+0.17}_{-0.21}$ & \textit{$\mathcal{U}$}(-1, 1) \\ [3pt]
	$\sqrt{e}\cdot\cos\omega$ & $-0.003^{+0.161}_{-0.155}$ & \textit{$\mathcal{U}$}(-1, 1) \\ [3pt]
	$a$ [au] &  $0.089\pm0.003$ & derived\\ [3pt]
	$m_{\rm p}\sini$ [$\mearth$] & $6.26^{+0.79}_{-0.76}$ & derived \\[15pt] 
	$RV_{\rm offset}$ [$\ms$] & $-0.27^{+0.49}_{-0.43}$ & \textit{$\mathcal{U}$}(-5, +5) \\ [3pt] 
	\hline 
	 & & \\ 
	 (hyper)Parameter   & Value &  Prior \\
	 (stellar activity)  & &   \\ [2pt]
	 \hline \\
	 $h$ [$\ms$]   &  $3.07^{+0.29}_{-0.24}$ & \textit{$\mathcal{U}$}(0,$+\infty$) \\ [3pt]
	 $\lambda$ [days] &   $34.4^{+11.6}_{-2.0}$  & \textit{$\mathcal{U}$}(0,$+\infty$) \\ [3pt]
	 $w$     &  $0.41^{+0.05}_{-0.04}$  & \textit{$\mathcal{U}$}(0,$+\infty$) \\ [3pt]
	 $\theta$ [days] &  $31.8^{+0.6}_{-0.5}$  & \textit{$\mathcal{U}$}($25, 45$)\\ [3pt]
	 $\sigma_{jitter}$ [$\ms$] &  $1.19^{+0.11}_{-0.14}$  &  \textit{$\mathcal{U}$}($0, 20$)\\\noalign{\smallskip}
	 \hline
	\end{tabular}
    \tablefoot{The symbol \textit{$\mathcal{U}$}($\cdot$,$\cdot$) denotes an uninformative prior with corresponding lower and upper limits.}
    \end{table*}

\section{Discussion and conclusions}\label{sec:concl}
We have presented in this paper the first results from the HADES programme conducted with HARPS-N at TNG. We have found a planetary system with two super-Earths hosted by the M1 dwarf GJ\,3998, by analysing the high precision, high resolution RV measurements in conjunction with almost-simultaneous photometric observations. We based our analysis on RV measurements obtained with the TERRA pipeline, which enables us to circumvent potential mismatch of the spectral lines that may happen using a CCF technique for M stars. 
The homogeneous analysis of the RV observations was carried out both with a frequentistic approach and by comparing models with a varying number of Keplerian signals and different models of the stellar activity noise.
The analysis of the radial velocity time series unveiled the presence of at least four significant periodic signals, two of these are linked to the activity of the host star and two to orbital periods:
\begin{itemize}
\item{P = 30.7 d, gives us an estimate of the rotational period of the star;} 
\item{P = 42.5 d, could be indicative of a modulation of the stellar variability due to differential rotation;} 
\item{P = 2.6 d, orbital period of GJ\,3998b;}
\item{P = 13.7 d, orbital period of GJ\,3998c.}
\end{itemize}
The conclusions on activity-related periods are confirmed by the analyses of the
activity indicators and of the photometric light curves of two independent sets of observations. The time series of $S$ index and H$\alpha$ show periodic variations around the 30.7 d and 42.5 d signals. The photometric results from both programs confirm the presence of variations in the period range 30 d -- 32 d, highlighting the strong connection of the long RV periods to chromospheric activity and to its rotational modulation. 
Due to its smaller amplitude and effect than the period of 30.7~d, the periodicity at 42.5~d could not be detected in the photometric time series, having a time coverage worst than that of the spectroscopic ones, too. For the same reason, the search for any robust correlation like $BV$ photometry vs. H$\alpha$ or $S$ indexes is unfortunately inconclusive.\\
Our results show that the detection of small planets in these data could be hampered by activity-induced signals. As a target of the HADES programme, GJ\,3998 is included in several statistical studies with the aim to scrutinize in deep the activity properties and rotations of M-dwarfs (Perger et al. 2016, submitted, Maldonado et al. 2016, submitted, Su{\'a}rez Mascare{\~n}o et al., Scandariato et al., in preparation). In particular, the accurate analysis of activity indicators performed by Su{\'a}rez Mascare{\~n}o et al. (in preparation), including also 6 HARPS (ESO) observations of GJ\,3998 acquired more than five years before the start of the HADES programme, provides an estimate of $30.8\,\pm2.5\, d$ for the rotational period and the hint of an activity cycle 500-600 d long.\\
In general, differential rotation is not yet fully understood, see e.g. the discussion in  \citet{kuk11}, therefore empirical data are then preferable. \citet{rei03} made an analysis of solar-type stars using an analysis 
of the shape of the lines in the Fourier domain. They considered a calibration of the ratio between the positions of the two first  zeros of the power spectrum of the line profile as a function of the ratio $\alpha$ between equatorial and polar rotation rate. They obtained an anticorrelation between differential rotation and equatorial velocity: differential rotation is mostly evident in slow rotators. Since GJ3998 is indeed a slow rotator, this might be indeed the case. However, Reiners \& Schmitt only considered solar-type stars and they do not show that we could see such trends in time-series of M dwarfs.

The most relevant data for us are then those provided by the analysis of the exquisitely sensitive time series obtained with Kepler. Using Kepler data, \citet{rei15} analysed an extremely wide datasets of stars ranging from solar-type stars to M-dwarfs and confirmed the relation found by Reiners \&  Schmitt between differential rotation and period. We may directly compare our result for GJ\,3998 with Figure 9 of Reinhold \& Gizon, where they plotted the minimum rotation period (that in our case is 30.7 d),
with the parameter $\alpha$=(Pmax-Pmin)/Pmax. For GJ\,3998, $\alpha$=(42.5-30.7)/42.5=0.277. The point for GJ\,3998 falls exactly in the middle
of the points for the M-stars. We would conclude that the assumption that the two periods related to activity for GJ\,3998 are due to rotation is fully consistent with what we know about differential rotation.

We run an MCMC simulation and use Bayesian model selection to determine the number of planets in the system and estimate their orbital parameters and minimum masses. We test several different models, varying the number of planets, eccentricity and treatment of stellar activity noise. We select a model involving two Keplerian signals, with a circular orbit for the inner planet and the eccentricity of the outer planet treated as a free parameter. The short-term stellar activity is modelled with the Gaussian Processes approach and the long-term activity noise is not included. 

\begin{figure}
\includegraphics[width=9cm]{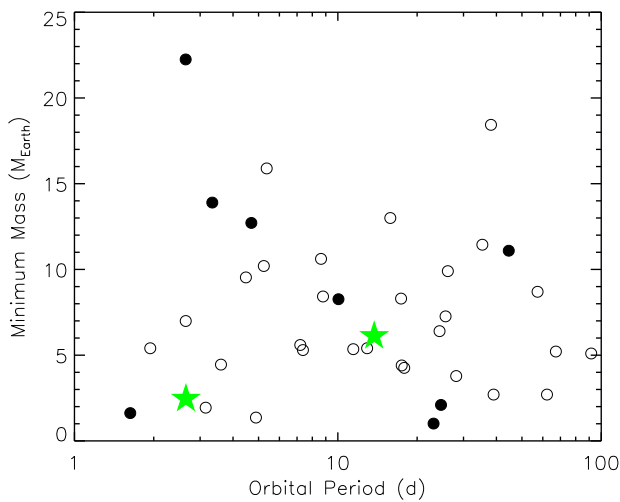}
\caption{Minimum mass vs. orbital period diagram for known Neptune-type and Super Earth planets around M dwarfs (http://www.exoplanet.eu - black dots, 2016 June 17), the green stars indicate GJ\,3998b and GJ\,3998c. Filled dots indicate planets with known radius.}
\label{fig:pianeti}
\end{figure}

The two planets appear to have minimum masses compatible with those of super-Earths, the inner planet has a minimum mass of $2.47 \pm\,0.27$\, M$_\oplus$, at a distance of 0.029 AU from the host star, the outer has a minimum mass of $6.26^{+0.79}_{-0.76}$\, M$_\oplus$ and a semi-major axis of 0.089 AU. \\A very rough estimate of the equilibrium temperatures of the planets, from the Stefan--Boltzmann law assuming uniform equilibrium
temperature over the entire planet and zero albedo, gives T$_{eq}$ = 740 K for the inner, and T$_{eq}$ = 420 K, for the outer planet. \\
These close distances strongly call for a search of potential transits in the next months, in particular for the inner planet, through a proposal for observing time on the $\it Spitzer\, Space\,
Telescope$. This planet has an interestingly high geometric transit probability of $\approx$\,8\%. With a minimal mass of $2.47 \pm\,0.27$\, M$_\oplus$, GJ\,3998b's radius should most likely lie somewhere between $\approx$\, 1 M$_\oplus$ (metal-rich composition) and $\approx$\, 1.65 M$_\oplus$ (ice-rich composition) \citep{sea07}. Given its host star's radius of $\approx$\, 0.5 R$_\odot$, this range of planetary sizes correspond to plausible transit dephts between 325 to 935 ppm. We thus proposed to use 20hr of $\it Spitzer$ time to monitor a 2-$\sigma$ transit window of the innermost planet in the GJ\,3998 system (P.I. M. Gillon).
A potential transit could provide a constraining point in the mass-radius diagram of known planets, enabling also the determination of the mean density and a better
characterization of the system architecture with future follow-up observations. In particular, there is currently no object with an accurately measured mass ($\le$\,20\%) in the range 2 -- 3 M$_\oplus$. This mass gap is even larger if we consider only M dwarf planets, as is evident in Fig. \ref{fig:pianeti}, which shows the minimum mass vs. orbital period diagram for known Neptune-type and Super Earth planets around main sequence M stars, along with the location of GJ\,3998 planets. The eventuality of a transit, given the small distance from the host and the brightness of GJ\,3998 (V = 10.83), would make this star a natural target for follow-up observations with the ESA $CHEOPS$ space telescope and also one of the most interesting M-dwarf targets for a detailed atmospheric characterization.

\begin{acknowledgements}
We thanks Guillem Anglada-Escud{\'e} for the distribution of the latest version of the TERRA pipeline. We are also grateful to Michael Gillon for helpful information on the search for transit of the inner planet with the $\it Spitzer\, Space\, Telescope$. \\ GAPS acknowledges support from INAF trough the "Progetti
Premiali'' funding scheme of the Italian Ministry of Education, University, and Research.\\
The research leading to these results has received funding from the European Union Seventh Framework Programme (FP7/2007-2013) under Grant Agreement No. 313014 (ETAEARTH).\\
L. A. acknowledges support from the Ariel ASI-INAF agreement N. 2015-038-R.0.\\
M. P., I. R., J. C. M., A. R., E. H, and M. L. acknowledge support from the Spanish Ministry of Economy and Competitiveness (MINECO) and the Fondo Europeo de Desarrollo Regional (FEDER) through grants ESP2013-48391-C4-1-R and ESP2014-57495-C2-2-R.\\
JIGH acknowledges financial support from the Spanish Ministry of Economy 
and Competitiveness (MINECO) under the 2013 Ram{\'o}n y Cajal program MINECO 
RYC-2013-14875, and ASM, JIGH, and RRL also acknowledge financial support 
from the Spanish ministry project MINECO AYA2014-56359-P. This article is based on observations made with the Carlos S\'anchez Telescope operated on the island of Tenerife by the IAC in the Spanish Observatorio del Teide.\\
We wish to thank the anonymous referee for thoughtful comments and suggestions that helped improve the manuscript.
\end{acknowledgements}

\bibliographystyle{aa}
\bibliography{Bibliographyrv}

\appendix
\section{Observing log for the GJ\,3998 spectra and results}
In this section we report the observing log for the GJ\,3998 spectra and the RVs, S and H$\alpha$ indices
we obtained with the present study. We list the observation dates (barycentric Julian date or BJD), the signal-to-noise ratios (S/Ns), the radial velocities (RVs) from the DRS and TERRA pipelines and the H$\alpha$ and $S$ indexes, calculated by the TERRA pipeline and in the present study.
The RV errors reported are the formal ones, not including the jitter term. The $S$ index and H$\alpha$ errors are calculated as described in the text and do not take into account the photon noise (Sect. \ref{sec:Prot}). The $S$ index and H$\alpha$ errors derived from the TERRA pipeline are due to photon noise through error propagation. 

 \onecolumn
 \scriptsize
\begin{center}
\begin{longtable}{cc|cc|cc|c |cc}
\caption{\label{Tab1} Data of 136 observed HARPS-N spectra of GJ\,3998. We list RVs obtained using the DRS and TERRA pipelines, $S$ index and H$\alpha$ calculated by the TERRA
pipeline (indicated with a 'T') and those derived in the present work, with an independent method.   }\\
 
  \hline \hline
BJD-2456000 & S/N & RV TERRA   & RV DRS   & $S$ index(T)   & H$\alpha$ index(T)    & $S$ index   & H$\alpha$\, index\\
(d) &   & (ms$^{-1}$)   & (ms$^{-1}$)   &    &    &  &  &\\ \hline 
 \endfirsthead
 \caption{continued.}\\
 
 \hline \hline
 BJD-2456000 & S/N & RV TERRA   & RV DRS   & $S$ index(T)   & H$\alpha$ index(T)    & $S$ index   & H$\alpha$ index \\
 (d) &   & (m $s^{-1}$)   & (m $s^{-1}$)   &    &     &  &  &\\ \hline 
 \endhead
\hline
 \endfoot
439.55832   & 54.1 & -3.661$\pm$0.959 & -44810.390$\pm$1.403 &  1.1013$\pm$0.0087 & 0.8357$\pm$0.0111 &    0.1025$\pm$0.0003  &  0.0591$\pm$0.0002 \\
439.65014   & 54.9 & -4.089$\pm$0.839 & -44810.869$\pm$1.372 &  1.0918$\pm$0.0085 & 0.8366$\pm$0.0106 &    0.0991$\pm$0.0004  &  0.0592$\pm$0.0001 \\
441.56648   & 56.9 &  3.221$\pm$0.945 & -44805.754$\pm$1.317 & 1.1114$\pm$0.0083 & 0.8370$\pm$0.0098 &    0.1026$\pm$0.0002  &  0.0593$\pm$0.0001 \\
442.54349   & 62.7 & -0.153$\pm$0.819 & -44806.802$\pm$1.185 &  1.1127$\pm$0.0075 & 0.8367$\pm$0.0086 &    0.1016$\pm$0.0003  &  0.0591$\pm$0.0001 \\
443.53644   & 36.0 &  2.606$\pm$1.316 & -44800.675$\pm$2.070 & 1.0772$\pm$0.0126 & 0.8348$\pm$0.0143 &    0.0982$\pm$0.0002  &  0.0589$\pm$0.0001 \\
443.63985   & 25.2 &  4.881$\pm$1.860 & -44797.699$\pm$2.907 & 1.1528$\pm$0.0179 & 0.8366$\pm$0.0183 &    0.1066$\pm$0.0007  &  0.0591$\pm$0.0001 \\
444.49963   & 54.4 &  9.699$\pm$0.779 & -44794.175$\pm$1.375 & 1.1731$\pm$0.0091 & 0.8348$\pm$0.0099 &    0.1050$\pm$0.0002  &  0.0589$\pm$0.0001 \\
444.72822   & 45.1 &  3.672$\pm$1.040 & -44803.153$\pm$1.659 & 1.1090$\pm$0.0106 & 0.8326$\pm$0.0118 &    0.1023$\pm$0.0003  &  0.0587$\pm$0.0001 \\
483.60939   & 18.3 & -2.992$\pm$2.562 & -44805.750$\pm$4.157 & 1.2120$\pm$0.0258 & 0.8477$\pm$0.0249 &    0.1046$\pm$0.0003  &  0.0595$\pm$0.0001 \\
485.59901   & 40.5 & -2.053$\pm$1.313 & -44810.630$\pm$1.823 &  1.2590$\pm$0.0129 & 0.8427$\pm$0.0117 &    0.1143$\pm$0.0004  &  0.0594$\pm$0.0001 \\
536.50393   & 43.7 & -0.322$\pm$1.182 & -44809.067$\pm$1.696 & 1.1779$\pm$0.0122 & 0.8610$\pm$0.0107 &    0.1110$\pm$0.0003  &  0.0605$\pm$0.0001 \\
693.77578   & 37.3 &  6.167$\pm$1.136 & -44800.357$\pm$1.907 & 1.2448$\pm$0.0138 & 0.8405$\pm$0.0116 &    0.1109$\pm$0.0001  &  0.0598$\pm$0.0001 \\
696.78604   & 27.5 &  0.727$\pm$1.817 & -44805.115$\pm$2.626 & 1.3229$\pm$0.0184 & 0.8442$\pm$0.0160 &    0.1205$\pm$0.0007  &  0.0600$\pm$0.0001 \\
697.76988   & 34.2 &  1.860$\pm$1.722 & -44806.957$\pm$2.162 & 1.2084$\pm$0.0148 & 0.8404$\pm$0.0139 &    0.1120$\pm$0.0005  &  0.0598$\pm$0.0001 \\
698.75906   & 31.5 &  2.978$\pm$1.142 & -44808.249$\pm$2.357 & 1.2340$\pm$0.0160 & 0.8368$\pm$0.0153 &    0.1186$\pm$0.0003  &  0.0593$\pm$0.0001 \\
699.73995   & 43.7 & -5.962$\pm$1.379 & -44816.402$\pm$1.667 &  1.2525$\pm$0.0123 & 0.8440$\pm$0.0107 &    0.1165$\pm$0.0002  &  0.0601$\pm$0.0001 \\
700.76184   & 48.7 & -2.033$\pm$1.096 & -44810.392$\pm$1.561 &  1.2810$\pm$0.0111 & 0.8452$\pm$0.0114 &    0.1147$\pm$0.0003  &  0.0603$\pm$0.0001 \\
701.72684   & 43.4 & -2.388$\pm$1.113 & -44811.385$\pm$1.705 &  1.2905$\pm$0.0121 & 0.8591$\pm$0.0113 &    0.1143$\pm$0.0003  &  0.0613$\pm$0.0001 \\
702.74632   & 38.5 & -5.141$\pm$1.050 & -44813.856$\pm$1.816 &  1.2109$\pm$0.0128 & 0.8475$\pm$0.0110 &    0.1097$\pm$0.0002  &  0.0604$\pm$0.0001 \\
751.75401   & 40.0 &  1.987$\pm$1.011 & -44805.977$\pm$1.815 & 1.4076$\pm$0.0144 & 0.8627$\pm$0.0140 &    0.1222$\pm$0.0006  &  0.0615$\pm$0.0001 \\
786.68698   & 25.0 &  0.206$\pm$1.864 & -44802.902$\pm$2.846 &  1.2481$\pm$0.0203 & 0.8643$\pm$0.0190 &    0.1132$\pm$0.0004  &  0.0613$\pm$0.0001 \\
787.67840   & 36.6 & -0.009$\pm$1.048 & -44811.380$\pm$1.967 &  1.3257$\pm$0.0151 & 0.8607$\pm$0.0146 &    0.1156$\pm$0.0006  &  0.0610$\pm$0.0002 \\
792.55372   & 35.2 & -8.221$\pm$1.142 & -44815.792$\pm$2.029 &  1.1524$\pm$0.0140 & 0.8522$\pm$0.0143 &    0.1061$\pm$0.0003  &  0.0604$\pm$0.0002 \\
811.61597   & 43.3 & -1.419$\pm$1.034 & -44810.187$\pm$1.603 &  1.2914$\pm$0.0124 & 0.8521$\pm$0.0108 &    0.1102$\pm$0.0004  &  0.0601$\pm$0.0001 \\
817.50088   & 57.4 &  7.990$\pm$0.728 & -44802.052$\pm$1.235 & 1.2861$\pm$0.0093 & 0.8547$\pm$0.0090 &    0.1141$\pm$0.0004  &  0.0603$\pm$0.0001 \\
818.51085   & 53.5 &  6.456$\pm$0.845 & -44803.281$\pm$1.335 & 1.3160$\pm$0.0100 & 0.8613$\pm$0.0099 &    0.1146$\pm$0.0003  &  0.0609$\pm$0.0001 \\
819.54712   & 42.3 &  1.268$\pm$0.937 & -44808.693$\pm$1.698 & 1.3058$\pm$0.0126 & 0.8706$\pm$0.0129 &    0.1139$\pm$0.0004  &  0.0617$\pm$0.0001 \\
820.60166   & 36.1 &  2.051$\pm$1.045 & -44805.066$\pm$1.944 & 1.3172$\pm$0.0144 & 0.8725$\pm$0.0136 &    0.1152$\pm$0.0001  &  0.0618$\pm$0.0001 \\
821.58470   & 33.7 & -1.340$\pm$1.146 & -44812.067$\pm$2.136 &  1.2561$\pm$0.0151 & 0.8712$\pm$0.0154 &    0.1091$\pm$0.0002  &  0.0617$\pm$0.0001 \\
854.47373   & 58.1 &  0.191$\pm$0.865 & -44806.984$\pm$1.243 &  1.3722$\pm$0.0095 & 0.8671$\pm$0.0101 &    0.1196$\pm$0.0004  &  0.0620$\pm$0.0001 \\
855.46144   & 48.3 & -0.165$\pm$0.943 & -44809.351$\pm$1.504 &  1.3440$\pm$0.0113 & 0.8645$\pm$0.0121 &    0.1184$\pm$0.0004  &  0.0620$\pm$0.0001 \\
857.51093   & 54.3 &  1.226$\pm$0.892 & -44807.754$\pm$1.326 & 1.3512$\pm$0.0102 & 0.8702$\pm$0.0105 &    0.1185$\pm$0.0003  &  0.0623$\pm$0.0002 \\
858.48015   & 45.8 & -3.806$\pm$0.826 & -44810.630$\pm$1.533 &  1.2590$\pm$0.0117 & 0.8530$\pm$0.0109 &    0.1091$\pm$0.0004  &  0.0608$\pm$0.0001 \\
859.48805   & 48.7 & -1.024$\pm$0.772 & -44807.930$\pm$1.434 &  1.2478$\pm$0.0110 & 0.8493$\pm$0.0101 &    0.1104$\pm$0.0003  &  0.0603$\pm$0.0001 \\
860.46642   & 62.2 &  1.170$\pm$0.770 & -44806.837$\pm$1.154 & 1.2115$\pm$0.0084 & 0.8463$\pm$0.0089 &    0.1101$\pm$0.0003  &  0.0601$\pm$0.0001 \\
861.47978   & 62.2 & -3.034$\pm$0.776 & -44811.458$\pm$1.167 &  1.1594$\pm$0.0082 & 0.8417$\pm$0.0096 &    0.1056$\pm$0.0003  &  0.0596$\pm$0.0001 \\
874.49707   & 56.1 & -0.508$\pm$0.831 & -44808.344$\pm$1.306 &  1.2786$\pm$0.0096 & 0.8430$\pm$0.0106 &    0.1118$\pm$0.0003  &  0.0597$\pm$0.0001 \\
875.43028   & 24.0 &  2.318$\pm$1.672 & -44808.228$\pm$2.994 & 1.2882$\pm$0.0214 & 0.8462$\pm$0.0198 &    0.1225$\pm$0.0003  &  0.0601$\pm$0.0001 \\
877.46015   & 50.1 & -0.246$\pm$0.789 & -44806.427$\pm$1.436 &  1.2887$\pm$0.0108 & 0.8583$\pm$0.0108 &    0.1152$\pm$0.0004  &  0.0608$\pm$0.0001 \\
878.46919   & 51.8 & -0.625$\pm$0.906 & -44807.814$\pm$1.409 &  1.3478$\pm$0.0108 & 0.8673$\pm$0.0114 &    0.1154$\pm$0.0002  &  0.0618$\pm$0.0002 \\
879.40909   & 60.5 & -1.048$\pm$0.891 & -44810.987$\pm$1.228 &  1.2844$\pm$0.0086 & 0.8628$\pm$0.0117 &    0.1134$\pm$0.0004  &  0.0615$\pm$0.0001 \\
880.40231   & 46.2 &  1.085$\pm$0.984 & -44808.277$\pm$1.575 & 1.2780$\pm$0.0118 & 0.8646$\pm$0.0124 &    0.1177$\pm$0.0002  &  0.0617$\pm$0.0001 \\
881.39121   & 26.7 &  1.366$\pm$1.390 & -44809.118$\pm$2.695 & 1.3753$\pm$0.0206 & 0.8622$\pm$0.0185 &    0.1278$\pm$0.0003  &  0.0612$\pm$0.0002 \\
892.38371   & 64.9 & -3.962$\pm$0.618 & -44811.950$\pm$1.122 &  1.1186$\pm$0.0076 & 0.8457$\pm$0.0090 &    0.1042$\pm$0.0004  &  0.0597$\pm$0.0001 \\
893.38677   & 50.3 & -2.915$\pm$0.884 & -44811.863$\pm$1.434 &  1.1567$\pm$0.0101 & 0.8407$\pm$0.0109 &    0.1057$\pm$0.0003  &  0.0594$\pm$0.0001 \\
894.38698   & 47.5 &  3.124$\pm$0.887 & -44805.304$\pm$1.516 & 1.1790$\pm$0.0110 & 0.8424$\pm$0.0115 &    0.1077$\pm$0.0003  &  0.0595$\pm$0.0001 \\
897.43710   & 43.8 &  7.581$\pm$1.051 & -44801.069$\pm$1.699 & 1.1314$\pm$0.0114 & 0.8453$\pm$0.0145 &    0.1058$\pm$0.0004  &  0.0595$\pm$0.0002 \\
898.39592   & 39.7 & -1.548$\pm$1.059 & -44810.103$\pm$1.833 &  1.1754$\pm$0.0128 & 0.8426$\pm$0.0138 &    0.1075$\pm$0.0005  &  0.0594$\pm$0.0001 \\
899.36885   & 40.9 &  2.599$\pm$0.949 & -44805.344$\pm$1.810 & 1.1490$\pm$0.0124 & 0.8340$\pm$0.0152 &    0.1092$\pm$0.0002  &  0.0588$\pm$0.0002 \\
905.37587   & 57.7 &  2.655$\pm$0.808 & -44807.049$\pm$1.277 & 1.1822$\pm$0.0089 & 0.8428$\pm$0.0108 &    0.1101$\pm$0.0003  &  0.0594$\pm$0.0001 \\
907.37031   & 54.0 &  3.483$\pm$0.806 & -44801.996$\pm$1.360 & 1.2569$\pm$0.0099 & 0.8464$\pm$0.0112 &    0.1126$\pm$0.0003  &  0.0598$\pm$0.0001 \\
909.40942   & 62.3 &  1.808$\pm$0.775 & -44807.678$\pm$1.195 & 1.2506$\pm$0.0084 & 0.8543$\pm$0.0099 &    0.1102$\pm$0.0004  &  0.0604$\pm$0.0001 \\
918.37437   & 41.5 &-10.356$\pm$0.951 & -44817.277$\pm$1.760 &   1.2339$\pm$0.0122 & 0.8493$\pm$0.0131 &    0.1114$\pm$0.0003  &  0.0597$\pm$0.0001\\
920.37594   & 57.6 & -7.259$\pm$0.743 & -44816.423$\pm$1.271 &  1.1552$\pm$0.0088 & 0.8463$\pm$0.0098 &    0.1041$\pm$0.0003  &  0.0596$\pm$0.0001 \\
928.34905   & 13.7 &  0.0  $\pm$3.377 & -44799.135$\pm$3.903 & 1.1199$\pm$0.0265 & 0.8449$\pm$0.0188 &    0.1156$\pm$0.0004  &  0.0603$\pm$0.0002 \\
932.35439   & 28.7 & -0.032$\pm$1.764 & -44809.760$\pm$2.564 &  1.1411$\pm$0.0171 & 0.8431$\pm$0.0189 &    0.1005$\pm$0.0004  &  0.0594$\pm$0.0001 \\
937.34656   & 58.6 &  5.878$\pm$0.761 & -44804.756$\pm$1.248 & 1.1774$\pm$0.0089 & 0.8452$\pm$0.0096 &    0.1069$\pm$0.0003  &  0.0594$\pm$0.0001 \\
938.34606   & 47.2 &  5.473$\pm$0.939 & -44803.136$\pm$1.539 & 1.1855$\pm$0.0113 & 0.8437$\pm$0.0114 &    0.1029$\pm$0.0003  &  0.0593$\pm$0.0001 \\
940.33204   & 56.1 &  4.844$\pm$0.947 & -44800.757$\pm$1.315 & 1.2549$\pm$0.0096 & 0.8559$\pm$0.0104 &    0.1156$\pm$0.0003  &  0.0603$\pm$0.0001 \\
942.33011   & 54.1 &  5.760$\pm$0.957 & -44803.151$\pm$1.363 & 1.2394$\pm$0.0098 & 0.8601$\pm$0.0107 &    0.1111$\pm$0.0002  &  0.0607$\pm$0.0001 \\
943.32501   & 30.7 &  1.592$\pm$1.181 & -44808.101$\pm$2.348 & 1.2875$\pm$0.0169 & 0.8566$\pm$0.0159 &    0.1144$\pm$0.0004  &  0.0604$\pm$0.0001 \\
1070.75888  & 30.2 & -9.021$\pm$1.229 & -44818.481$\pm$2.258 &  0.9596$\pm$0.0157 & 0.8328$\pm$0.0149 &    0.0989$\pm$0.0001  &  0.0588$\pm$0.0001 \\
1090.74239  & 32.9 &  5.923$\pm$1.059 & -44800.869$\pm$1.975 & 1.1940$\pm$0.0148 & 0.8281$\pm$0.0121 &    0.1104$\pm$0.0004  &  0.0582$\pm$0.0001 \\
1091.72752  & 34.6 & -0.157$\pm$1.036 & -44809.882$\pm$1.941 &  1.1686$\pm$0.0143 & 0.8359$\pm$0.0125 &    0.1059$\pm$0.0002  &  0.0591$\pm$0.0001 \\
1091.76300  & 40.0 &  0.016$\pm$0.827 & -44807.609$\pm$1.684 &  1.1434$\pm$0.0123 & 0.8282$\pm$0.0109 &    0.1025$\pm$0.0003  &  0.0584$\pm$0.0001 \\
1092.73829  & 39.5 & -1.897$\pm$0.889 & -44806.608$\pm$1.696 &  1.1975$\pm$0.0127 & 0.8373$\pm$0.0110 &    0.1100$\pm$0.0005  &  0.0592$\pm$0.0001 \\
1093.72821  & 46.8 &  0.560$\pm$0.930 & -44809.135$\pm$1.498 & 1.1893$\pm$0.0111 & 0.8322$\pm$0.0109 &    0.1106$\pm$0.0003  &  0.0588$\pm$0.0001 \\
1094.73327  & 40.5 & -6.982$\pm$1.084 & -44816.980$\pm$1.669 &  1.1873$\pm$0.0124 & 0.8335$\pm$0.0109 &    0.1097$\pm$0.0004  &  0.0588$\pm$0.0001 \\
1096.75497  & 28.0 &-10.235$\pm$1.392 & -44815.323$\pm$2.473 &   1.1516$\pm$0.0173 & 0.8336$\pm$0.0165 &    0.0987$\pm$0.0003  &  0.0586$\pm$0.0001\\
1097.72784  & 46.0 &-10.425$\pm$0.883 & -44819.188$\pm$1.529 &   1.1253$\pm$0.0107 & 0.8355$\pm$0.0113 &    0.1057$\pm$0.0004  &  0.0589$\pm$0.0001\\
1098.72675  & 48.6 & -7.595$\pm$0.832 & -44817.919$\pm$1.439 &  1.1110$\pm$0.0100 & 0.8412$\pm$0.0105 &    0.1015$\pm$0.0003  &  0.0595$\pm$0.0001 \\
1099.72238  & 30.5 & -6.790$\pm$1.508 & -44818.442$\pm$2.252 &  1.0209$\pm$0.0150 & 0.8400$\pm$0.0152 &    0.0937$\pm$0.0004  &  0.0594$\pm$0.0001 \\
1108.72655  & 39.4 & -0.184$\pm$1.021 & -44808.811$\pm$1.739 &  1.1809$\pm$0.0129 & 0.8432$\pm$0.0119 &    0.1032$\pm$0.0003  &  0.0593$\pm$0.0001 \\
1109.68659  & 23.8 & -0.005$\pm$1.802 & -44804.934$\pm$2.817 &  1.0690$\pm$0.0191 & 0.8328$\pm$0.0174 &    0.0909$\pm$0.0002  &  0.0587$\pm$0.0001 \\
1112.64273  & 25.2 &  1.303$\pm$1.764 & -44808.052$\pm$2.879 & 1.0746$\pm$0.0181 & 0.8333$\pm$0.0215 &    0.1144$\pm$0.0004  &  0.0590$\pm$0.0001 \\
1113.63422  & 35.7 & -4.573$\pm$1.052 & -44815.389$\pm$2.004 &  1.2025$\pm$0.0145 & 0.8367$\pm$0.0152 &    0.1066$\pm$0.0005  &  0.0590$\pm$0.0001 \\
1114.68664  & 11.5 &  1.580$\pm$3.282 & -44811.796$\pm$6.380 & 0.8582$\pm$0.0284 & 0.8370$\pm$0.0347 &    0.1159$\pm$0.0025  &  0.0588$\pm$0.0001 \\
1139.69050  & 39.0 & -6.536$\pm$1.015 & -44815.403$\pm$1.800 &  1.1410$\pm$0.0126 & 0.8435$\pm$0.0127 &    0.1043$\pm$0.0003  &  0.0593$\pm$0.0001 \\
1140.60385  & 24.7 & -3.590$\pm$1.312 & -44815.904$\pm$2.722 &  1.0263$\pm$0.0175 & 0.8387$\pm$0.0165 &    0.0914$\pm$0.0002  &  0.0591$\pm$0.0001 \\
1142.69371  & 51.3 & -2.458$\pm$0.932 & -44810.821$\pm$1.409 &  1.2382$\pm$0.0100 & 0.8449$\pm$0.0120 &    0.1122$\pm$0.0004  &  0.0598$\pm$0.0001 \\
1143.64481  & 43.1 &  1.599$\pm$1.049 & -44808.220$\pm$1.650 & 1.1056$\pm$0.0110 & 0.8412$\pm$0.0128 &    0.1000$\pm$0.0003  &  0.0591$\pm$0.0001 \\
1144.63786  & 41.3 &  2.201$\pm$0.850 & -44806.596$\pm$1.735 & 1.0223$\pm$0.0111 & 0.8367$\pm$0.0140 &    0.0966$\pm$0.0004  &  0.0588$\pm$0.0001 \\
1145.66091  & 28.3 &  2.836$\pm$1.706 & -44807.460$\pm$2.576 & 0.9863$\pm$0.0154 & 0.8496$\pm$0.0188 &    0.0866$\pm$0.0003  &  0.0597$\pm$0.0002 \\
1146.67150  & 43.5 &  5.777$\pm$0.990 & -44802.194$\pm$1.562 & 1.1015$\pm$0.0110 & 0.8417$\pm$0.0103 &    0.1077$\pm$0.0005  &  0.0594$\pm$0.0002 \\
1147.70960  & 44.3 &  2.616$\pm$0.803 & -44803.476$\pm$1.562 & 1.1917$\pm$0.0112 & 0.8406$\pm$0.0106 &    0.1066$\pm$0.0005  &  0.0592$\pm$0.0002 \\
1148.72267  & 54.0 &  1.324$\pm$0.780 & -44808.025$\pm$1.314 & 1.1911$\pm$0.0094 & 0.8467$\pm$0.0097 &    0.1066$\pm$0.0004  &  0.0597$\pm$0.0002 \\
1150.59928  & 34.1 & -0.044$\pm$1.316 & -44804.505$\pm$2.047 &  1.1567$\pm$0.0143 & 0.8445$\pm$0.0139 &    0.1057$\pm$0.0004  &  0.0593$\pm$0.0001 \\
1153.50327  & 22.6 & -1.839$\pm$1.866 & -44812.419$\pm$3.140 &  1.2529$\pm$0.0225 & 0.8479$\pm$0.0205 &    0.1197$\pm$0.0005  &  0.0599$\pm$0.0001 \\
1156.53267  & 24.6 & -0.769$\pm$1.424 & -44808.516$\pm$2.531 &  1.2758$\pm$0.0194 & 0.8614$\pm$0.0144 &    0.1129$\pm$0.0002  &  0.0609$\pm$0.0002 \\
1159.52968  & 41.0 &  5.064$\pm$1.046 & -44803.803$\pm$1.772 & 1.2824$\pm$0.0130 & 0.8505$\pm$0.0147 &    0.1148$\pm$0.0003  &  0.0601$\pm$0.0002 \\
1166.53149  & 46.3 & -7.144$\pm$0.841 & -44815.420$\pm$1.511 &  1.3054$\pm$0.0115 & 0.8445$\pm$0.0109 &    0.1186$\pm$0.0004  &  0.0601$\pm$0.0001 \\
1169.52737  & 53.1 & -5.950$\pm$0.891 & -44812.443$\pm$1.326 &  1.1881$\pm$0.0095 & 0.8331$\pm$0.0102 &    0.1062$\pm$0.0003  &  0.0590$\pm$0.0002 \\
1170.53362  & 55.7 & -4.948$\pm$0.757 & -44813.632$\pm$1.242 &  1.1707$\pm$0.0088 & 0.8314$\pm$0.0089 &    0.1054$\pm$0.0003  &  0.0588$\pm$0.0001 \\
1172.61490  & 45.6 & -3.195$\pm$0.924 & -44810.884$\pm$1.490 &  1.1263$\pm$0.0107 & 0.8293$\pm$0.0103 &    0.1029$\pm$0.0003  &  0.0584$\pm$0.0001 \\
1173.61586  & 39.4 & -2.179$\pm$1.084 & -44810.531$\pm$1.741 &  1.0962$\pm$0.0122 & 0.8289$\pm$0.0125 &    0.1016$\pm$0.0002  &  0.0585$\pm$0.0001 \\
1174.48447  & 28.8 & -2.554$\pm$1.026 & -44812.054$\pm$2.393 &  1.0362$\pm$0.0155 & 0.8262$\pm$0.0164 &    0.1035$\pm$0.0004  &  0.0582$\pm$0.0001 \\
1175.54539  & 37.7 &  1.869$\pm$1.048 & -44804.167$\pm$1.861 & 1.1412$\pm$0.0129 & 0.8375$\pm$0.0142 &    0.1093$\pm$0.0005  &  0.0593$\pm$0.0001 \\
1178.59796  & 27.8 & -5.159$\pm$1.583 & -44810.597$\pm$2.446 &  1.1398$\pm$0.0166 & 0.8354$\pm$0.0162 &    0.1019$\pm$0.0001  &  0.0588$\pm$0.0001 \\
1203.51926  & 38.2 & -2.140$\pm$0.987 & -44810.314$\pm$1.798 &  1.1800$\pm$0.0130 & 0.8444$\pm$0.0125 &    0.1065$\pm$0.0004  &  0.0599$\pm$0.0001 \\
1203.52964  & 35.7 & -1.973$\pm$1.185 & -44807.610$\pm$1.888 &  1.1531$\pm$0.0136 & 0.8420$\pm$0.0126 &    0.1085$\pm$0.0002  &  0.0598$\pm$0.0001 \\
1204.51742  & 54.9 &  0.575$\pm$0.781 & -44805.036$\pm$1.244 & 1.2053$\pm$0.0091 & 0.8412$\pm$0.0087 &    0.1079$\pm$0.0003  &  0.0596$\pm$0.0001 \\
1205.49679  & 51.1 & -3.729$\pm$0.887 & -44810.929$\pm$1.384 &  1.1680$\pm$0.0099 & 0.8422$\pm$0.0109 &    0.1061$\pm$0.0003  &  0.0598$\pm$0.0001 \\
1206.49746  & 49.7 & -0.561$\pm$1.060 & -44808.114$\pm$1.427 &  1.2155$\pm$0.0100 & 0.8479$\pm$0.0121 &    0.1098$\pm$0.0003  &  0.0603$\pm$0.0001 \\
1207.51090  & 47.2 & -2.575$\pm$0.995 & -44810.004$\pm$1.494 &  1.2622$\pm$0.0110 & 0.8481$\pm$0.0119 &    0.1123$\pm$0.0004  &  0.0603$\pm$0.0001 \\
1208.50352  & 37.6 & -5.019$\pm$1.095 & -44809.261$\pm$1.807 &  1.2270$\pm$0.0134 & 0.8458$\pm$0.0143 &    0.1074$\pm$0.0002  &  0.0600$\pm$0.0001 \\
1209.57184  & 33.8 &  2.672$\pm$1.374 & -44803.759$\pm$2.056 & 1.3323$\pm$0.0160 & 0.8484$\pm$0.0144 &    0.1246$\pm$0.0006  &  0.0602$\pm$0.0001 \\
1210.56713  & 32.2 &  1.057$\pm$1.526 & -44804.398$\pm$2.150 & 1.4216$\pm$0.0176 & 0.8509$\pm$0.0147 &    0.1207$\pm$0.0002  &  0.0608$\pm$0.0001 \\
1239.43870  & 52.9 &  6.150$\pm$0.982 & -44800.882$\pm$1.349 & 1.2292$\pm$0.0097 & 0.8411$\pm$0.0105 &    0.1102$\pm$0.0003  &  0.0592$\pm$0.0001 \\
1240.43870  & 50.8 &  5.485$\pm$0.926 & -44801.580$\pm$1.418 & 1.2648$\pm$0.0103 & 0.8467$\pm$0.0116 &    0.1123$\pm$0.0004  &  0.0600$\pm$0.0001 \\
1241.43527  & 38.9 &  8.960$\pm$1.434 & -44799.033$\pm$1.840 & 1.2080$\pm$0.0129 & 0.8492$\pm$0.0140 &    0.1067$\pm$0.0003  &  0.0600$\pm$0.0001 \\
1242.43572  & 36.5 &  6.373$\pm$1.140 & -44799.945$\pm$1.974 & 1.3049$\pm$0.0144 & 0.8646$\pm$0.0152 &    0.1168$\pm$0.0005  &  0.0615$\pm$0.0001 \\
1249.46629  & 34.2 &  4.094$\pm$1.220 & -44802.309$\pm$2.082 & 1.4305$\pm$0.0163 & 0.8613$\pm$0.0148 &    0.1223$\pm$0.0004  &  0.0612$\pm$0.0001 \\
1250.44664  & 42.2 &  0.647$\pm$1.114 & -44808.952$\pm$1.758 & 1.3660$\pm$0.0124 & 0.8634$\pm$0.0171 &    0.1244$\pm$0.0002  &  0.0614$\pm$0.0001 \\
1251.42295  & 50.6 & -0.138$\pm$0.748 & -44810.502$\pm$1.447 &  1.4567$\pm$0.0111 & 0.8845$\pm$0.0123 &    0.1247$\pm$0.0004  &  0.0636$\pm$0.0001 \\
1259.40267  & 47.3 & -3.876$\pm$1.056 & -44812.051$\pm$1.494 &  1.2697$\pm$0.0113 & 0.8453$\pm$0.0109 &    0.1149$\pm$0.0004  &  0.0597$\pm$0.0001 \\
1260.42448  & 57.3 & -3.050$\pm$0.815 & -44812.072$\pm$1.270 &  1.1387$\pm$0.0085 & 0.8413$\pm$0.0104 &    0.1058$\pm$0.0003  &  0.0592$\pm$0.0001 \\
1261.40688  & 49.7 & -6.275$\pm$0.909 & -44813.246$\pm$1.449 &  1.1761$\pm$0.0101 & 0.8446$\pm$0.0115 &    0.1060$\pm$0.0004  &  0.0596$\pm$0.0001 \\
1262.40283  & 45.9 & -5.124$\pm$1.009 & -44815.320$\pm$1.569 &  1.2067$\pm$0.0110 & 0.8404$\pm$0.0123 &    0.1088$\pm$0.0003  &  0.0592$\pm$0.0001 \\
1263.40319  & 53.1 & -5.161$\pm$1.035 & -44812.106$\pm$1.368 &  1.1613$\pm$0.0092 & 0.8458$\pm$0.0114 &    0.1040$\pm$0.0002  &  0.0597$\pm$0.0001 \\
1264.40216  & 41.3 & -1.994$\pm$0.930 & -44809.384$\pm$1.763 &  1.1439$\pm$0.0116 & 0.8402$\pm$0.0150 &    0.1053$\pm$0.0002  &  0.0592$\pm$0.0001 \\
1274.38523  & 45.8 &  3.116$\pm$0.950 & -44803.807$\pm$1.516 & 1.2575$\pm$0.0112 & 0.8481$\pm$0.0103 &    0.1099$\pm$0.0003  &  0.0597$\pm$0.0001 \\
1275.38502  & 42.7 &  1.220$\pm$1.001 & -44804.871$\pm$1.676 & 1.2876$\pm$0.0124 & 0.8517$\pm$0.0123 &    0.1157$\pm$0.0005  &  0.0602$\pm$0.0001 \\
1276.38362  & 48.8 &  6.410$\pm$1.019 & -44800.954$\pm$1.476 & 1.2586$\pm$0.0107 & 0.8511$\pm$0.0111 &    0.1135$\pm$0.0005  &  0.0599$\pm$0.0001 \\
1277.38152  & 54.8 &  2.589$\pm$0.734 & -44805.118$\pm$1.323 & 1.3025$\pm$0.0097 & 0.8549$\pm$0.0101 &    0.1133$\pm$0.0002  &  0.0603$\pm$0.0001 \\
1278.41821  & 55.0 &  4.320$\pm$0.939 & -44803.871$\pm$1.337 & 1.3122$\pm$0.0098 & 0.8559$\pm$0.0109 &    0.1156$\pm$0.0003  &  0.0607$\pm$0.0001 \\
1282.39647  & 54.6 &  5.234$\pm$0.775 & -44801.250$\pm$1.334 & 1.4052$\pm$0.0103 & 0.8623$\pm$0.0103 &    0.1192$\pm$0.0004  &  0.0612$\pm$0.0001 \\
1285.39851  & 33.1 &  3.499$\pm$1.388 & -44805.311$\pm$2.160 & 1.3546$\pm$0.0169 & 0.8515$\pm$0.0150 &    0.1204$\pm$0.0003  &  0.0599$\pm$0.0001 \\
1286.38285  & 47.1 &  2.847$\pm$0.985 & -44805.138$\pm$1.542 & 1.2665$\pm$0.0112 & 0.8496$\pm$0.0118 &    0.1148$\pm$0.0004  &  0.0601$\pm$0.0001 \\
1287.38443  & 43.5 &  1.567$\pm$0.990 & -44806.769$\pm$1.651 & 1.2075$\pm$0.0115 & 0.8476$\pm$0.0120 &    0.1055$\pm$0.0003  &  0.0596$\pm$0.0001 \\
1289.41433  & 22.9 &  0.841$\pm$1.891 & -44806.672$\pm$3.147 & 1.1718$\pm$0.0211 & 0.8553$\pm$0.0200 &    0.1206$\pm$0.0003  &  0.0601$\pm$0.0002 \\
1291.42784  & 47.4 & -5.261$\pm$1.010 & -44813.194$\pm$1.521 &  1.1612$\pm$0.0113 & 0.8487$\pm$0.0108 &    0.1041$\pm$0.0005  &  0.0598$\pm$0.0001 \\
1293.38288  & 34.0 & -8.067$\pm$1.253 & -44816.542$\pm$2.101 &  1.1373$\pm$0.0144 & 0.8421$\pm$0.0146 &    0.0990$\pm$0.0002  &  0.0591$\pm$0.0001 \\
1297.38013  & 26.9 &  5.295$\pm$1.625 & -44803.143$\pm$2.287 & 1.0471$\pm$0.0167 & 0.8452$\pm$0.0128 &    0.0944$\pm$0.0004  &  0.0592$\pm$0.0001 \\
1306.34245  & 44.4 & -0.482$\pm$0.981 & -44806.765$\pm$1.657 &  1.2165$\pm$0.0117 & 0.8527$\pm$0.0143 &    0.1075$\pm$0.0001  &  0.0600$\pm$0.0001 \\
1307.33712  & 51.4 &  1.786$\pm$0.792 & -44803.888$\pm$1.414 & 1.3772$\pm$0.0108 & 0.8726$\pm$0.0114 &    0.1190$\pm$0.0003  &  0.0619$\pm$0.0002 \\
1308.33682  & 50.5 &  4.989$\pm$0.715 & -44801.524$\pm$1.429 & 1.2736$\pm$0.0105 & 0.8552$\pm$0.0109 &    0.1143$\pm$0.0003  &  0.0600$\pm$0.0001 \\
\hline \hline
\end{longtable}
\end{center}

\end{document}